\documentclass[aps,english,preprint,nofootinbib,preprintnumbers]{revtex4}
\usepackage{mathrsfs}
\usepackage{hyperref}
\hypersetup{
    colorlinks=true,
    linkcolor=blue,
    filecolor=magenta,      
    urlcolor=cyan,
    pdftitle={Overleaf Example},
    pdfpagemode=FullScreen,
    }
\usepackage{graphicx}
\usepackage{amsmath, amssymb}
\usepackage{babel}
\usepackage{color}
\usepackage{slashed}
\usepackage[T1]{fontenc}
\usepackage{titlesec}
\def\beqn{\begin{eqnarray}}
\def\eeqn{\end{eqnarray}}
\def\barr{\begin{array}}
\def\earr{\end{array}}
\def\btab{\begin{tabular}}
\def\etab{\end{tabular}}
\def\bite{\begin{itemize}}
\def\eite{\end{itemize}}
\def\bcen{\begin{center}}
\def\ecen{\end{center}}

\newcommand{\q}{\mathbf{q}}

\begin{document}
%\preprint{MITP/20-019}

\title{Dispersive formalism for the nuclear structure correction $\delta_\mathrm{NS}$ to the $\beta$ decay rate }

\author{Chien-Yeah Seng$^{1,2,3}$}
\author{Mikhail Gorchtein$^{4,5}$}

\affiliation{$^{1}$Helmholtz-Institut f\"{u}r Strahlen- und Kernphysik and Bethe Center for
	Theoretical Physics,\\ Universit\"{a}t Bonn, 53115 Bonn, Germany}
\affiliation{$^{2}$Facility for Rare Isotope Beams, Michigan State University, East Lansing, MI 48824, USA}
\affiliation{$^{3}$Department of Physics, University of Washington,
	Seattle, WA 98195-1560, USA}
\affiliation{$^{4}$Institut f\"ur Kernphysik, Johannes Gutenberg-Universit\"{a}t,\\
	J.J. Becher-Weg 45, 55128 Mainz, Germany}
\affiliation{$^{5}$PRISMA Cluster of Excellence, Johannes Gutenberg-Universit\"{a}t, Mainz, Germany}

\date{\today}

\begin{abstract}

We analyze the axial $\gamma W$-box diagram for $I(J^P)=1(0^+)$ nuclei and provide a dispersion representation of the nuclear-structure correction $\delta_\text{NS}$ including its energy-dependent part. We also summarize useful isospin rotation formula and representations in nuclear theory that could facilitate the calculation of the parity-odd nuclear structure function $F_3(\nu,Q^2)$. They provide a rigorous theory framework for the future, high-precision calculation of the nuclear structure correction $\delta_\text{NS}$ necessary for the extraction of the Cabibbo-Kobayashi-Maskawa matrix element $|V_{ud}|$ from superallowed nuclear $\beta$ decays.

\end{abstract}

\maketitle

  \tableofcontents

\section{Introduction}

The inability of the Standard Model (SM) of particle physics to explain phenomena such as dark energy, dark matter and matter-antimatter asymmetry indicates that it is an incomplete theory and must be extended. Small-scale experiments at the precision frontier play an important role in the search of physics beyond the Standard Model (BSM), and several interesting anomalies have been identified in, e.g., the measurement of the muon anomalous magnetic moment~\cite{Fermigm2,Aoyama:2020ynm,Miller:2007kk,Miller:2012opa,Jegerlehner:2009ry}, decays of B-mesons~\cite{Aaij:2019wad,Aaij:2014ora,Aaij:2015yra,Aaij:2015oid} and, more recently, the ``Cabibbo angle anomaly''. The latter consists in the mutual disagreement between different experiments in the determination of the Cabibbo angle $\theta_C$ from the first-row Cabibbo-Kobayashi-Maskawa (CKM) matrix elements $V_{ud}=\cos\theta_C$ and $V_{us}=\sin\theta_C$, and provides hints to possible solutions in terms of BSM physics~\cite{Crivellin:2020lzu,Crivellin:2021njn}. 

The past few years have seen a tremendous amount of progress, both in theory and experiment, relevant to the extraction of $V_{ud}$ and $V_{us}$ from charged weak decay processes. In the pion sector, a new lattice calculation~\cite{Feng:2020zdc} reduces the theory uncertainty in the $\pi^+\rightarrow \pi^0 e^+\nu$ decay and motivates the design of a future rare pion decay experiment PIONEER~\cite{PIONEER:2022alm,PIONEER:2022yag}. In the neutron sector, progress has been achieved in the theory calculation of the single-nucleon inner radiative correction (RC)~\cite{Seng:2018yzq,Seng:2018qru,Czarnecki:2019mwq,Seng:2020wjq,Hayen:2020cxh,Shiells:2020fqp,Gorchtein:2021fce,Cirigliano:2022hob} and experimental measurements of the neutron lifetime~\cite{Serebrov:2017bzo,Pattie:2017vsj,Ezhov:2014tna,UCNt:2021pcg} and axial coupling constant $g_A$~\cite{UCNA:2017obv,Markisch:2018ndu,Beck:2019xye,Hassan:2020hrj}. Improved lattice calculations of the bare-QCD axial coupling constant open a new window for BSM searches through the comparison between theory and experiment~\cite{Chang:2018uxx,Gupta:2018qil,Walker-Loud:2019cif}. In the kaon sector, theory improvements include new evaluation of the $K\rightarrow \pi\ell\nu$ RC~\cite{Seng:2019lxf,Seng:2020jtz,Ma:2021azh,Seng:2021boy,Seng:2021wcf,Seng:2021nar,Seng:2022wcw},  new lattice calculations of the $K\rightarrow\pi$ transition form factor~\cite{Bazavov:2018kjg,Kakazu:2019ltq,Ishikawa:2022otj} and the $K\rightarrow\mu\nu$ RC~\cite{Giusti:2017dwk,Giusti:2018guw}.
Experimental progress entails the first measurement of the $K_S\rightarrow\pi\mu\nu$ branching ratio~\cite{KLOE-2:2019rev}. Many of the recent developments serve to reduce the existing uncertainties in $V_{ud}$ and $V_{us}$, but some happen to increase them; an example of this kind occurs in the nuclear sector which is the focus of this work.

Superallowed $\beta$ decays of $J^P=0^+$ nuclei currently provide the most precise measurement of $V_{ud}$ through the following master formula: 
\begin{equation}
|V_{ud}|^2_{0^+}=\frac{2984.43~\text{s}}{\mathcal{F}t(1+\Delta_R^V)}~,\label{eq:Vudmaster}
\end{equation} 
where $\Delta_R^V$ is the same nucleus-independent inner RC as in free neutron. Meanwhile, the quantity $\mathcal{F}t$ combines the experimentally measured $ft$-values with all nucleus-dependent corrections to give another universal quantity,
\begin{equation}
\mathcal{F}t=ft(1+\delta_\text{R}')(1-\delta_\text{C}+\delta_\text{NS})~.\label{eq:Ftdef}
\end{equation}
Among the various corrections, $\delta_\text{R}'$ is the electron energy-dependent ``outer'' correction which was calculated to $Z^2\alpha^3$ and is well under control~\cite{Sirlin:1967zza,Sirlin:1987sy,Sirlin:1986cc}. In turn, $\delta_\text{C}$ is the isospin-symmetry breaking (ISB) correction to the Fermi matrix element, which has long been subject to intense debate~\cite{Towner:2002rg,Towner:2007np,Hardy:2008gy,Xayavong:2017kim,Ormand:1989hm,Ormand:1995df,Satula:2011br,Satula:2016hbs,Liang:2009pf,Auerbach:2008ut,Miller:2008my,Miller:2009cg,Condren:2022dji}. In this paper we focus on the last correction $\delta_\text{NS}$, the nuclear-structure-dependent part of the RC to superallowed $\beta$ decays.

\begin{figure}[tb]
	\begin{centering}
		\includegraphics[scale=0.15]{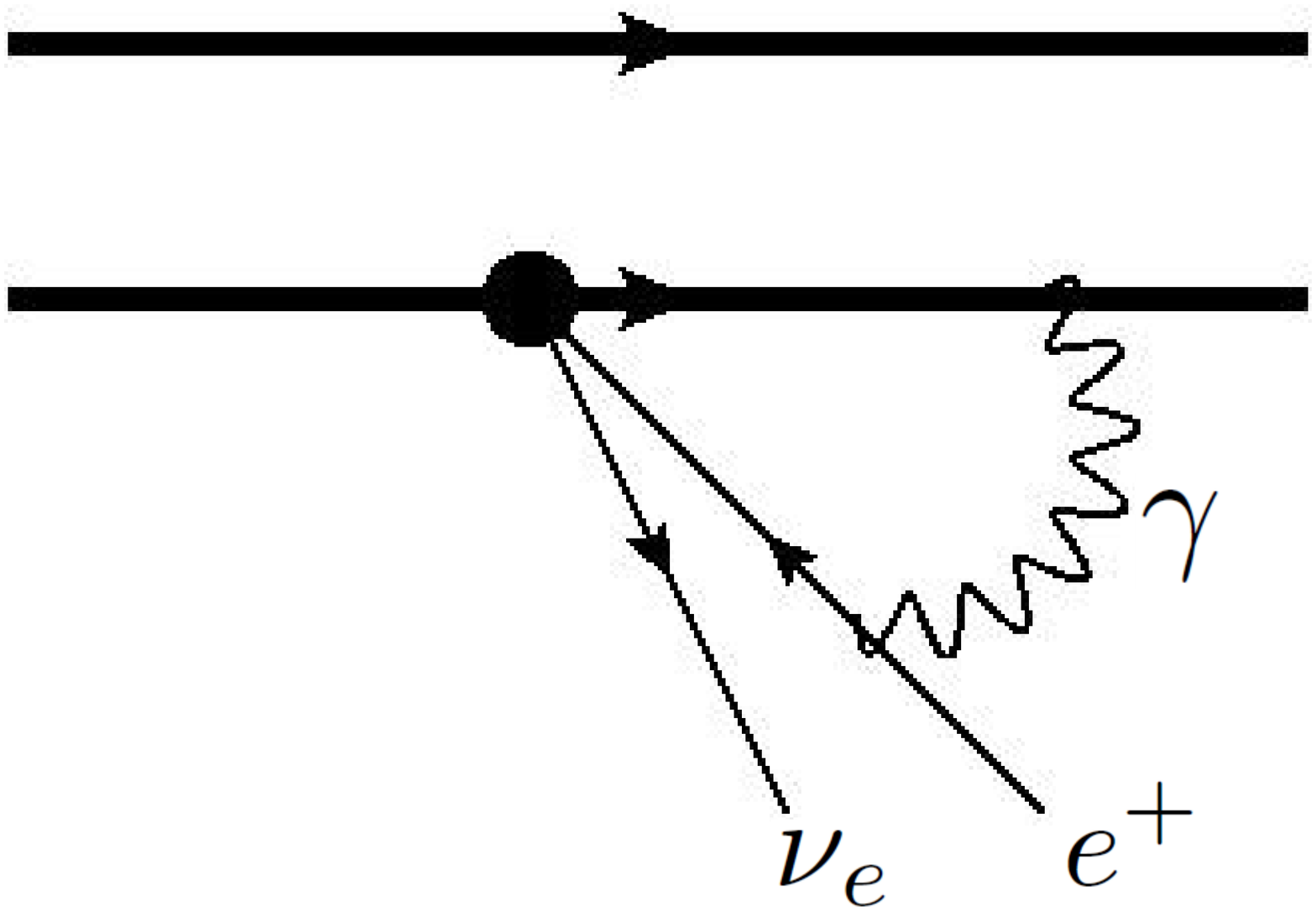}
		\includegraphics[scale=0.15]{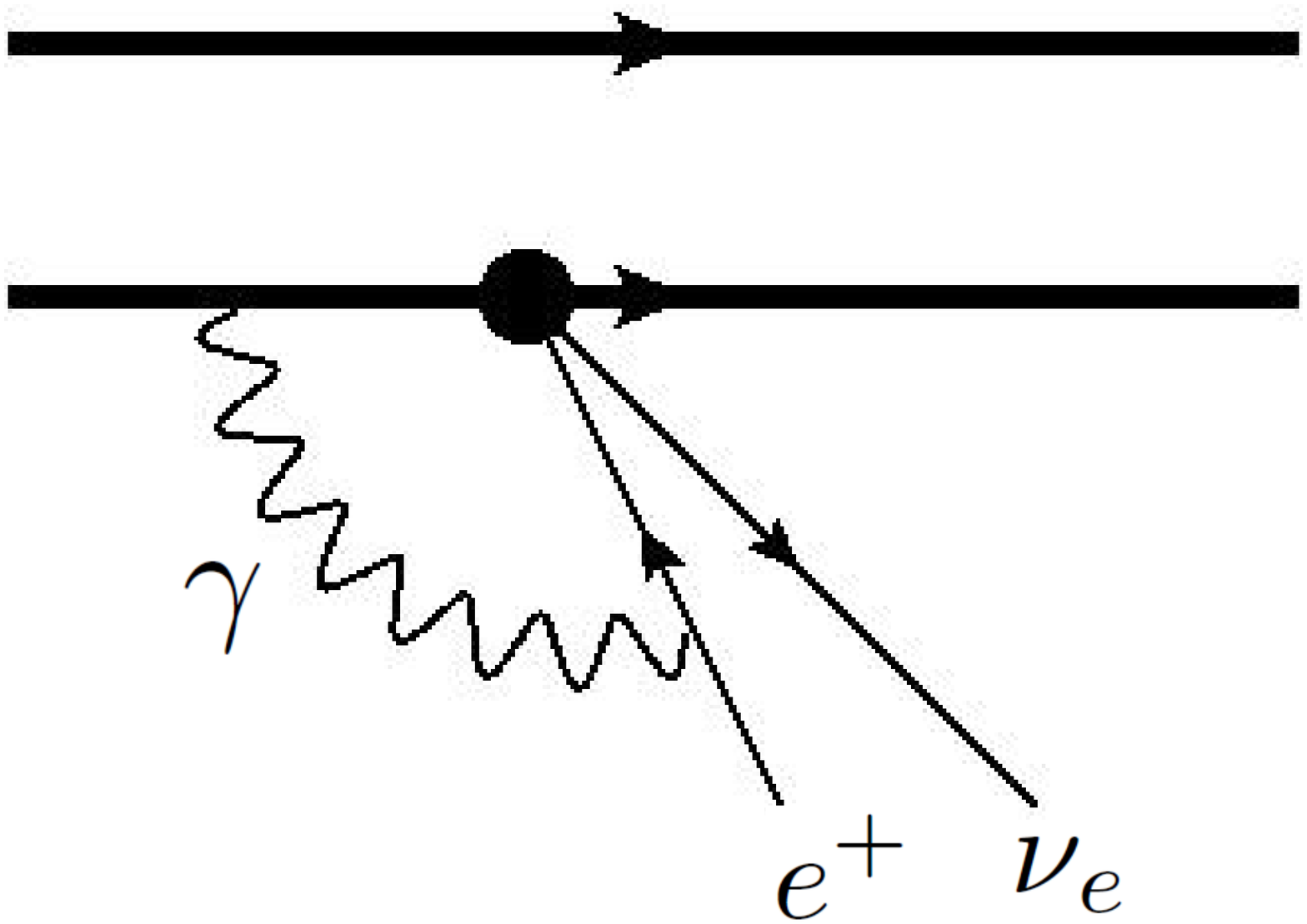}
		\includegraphics[scale=0.15]{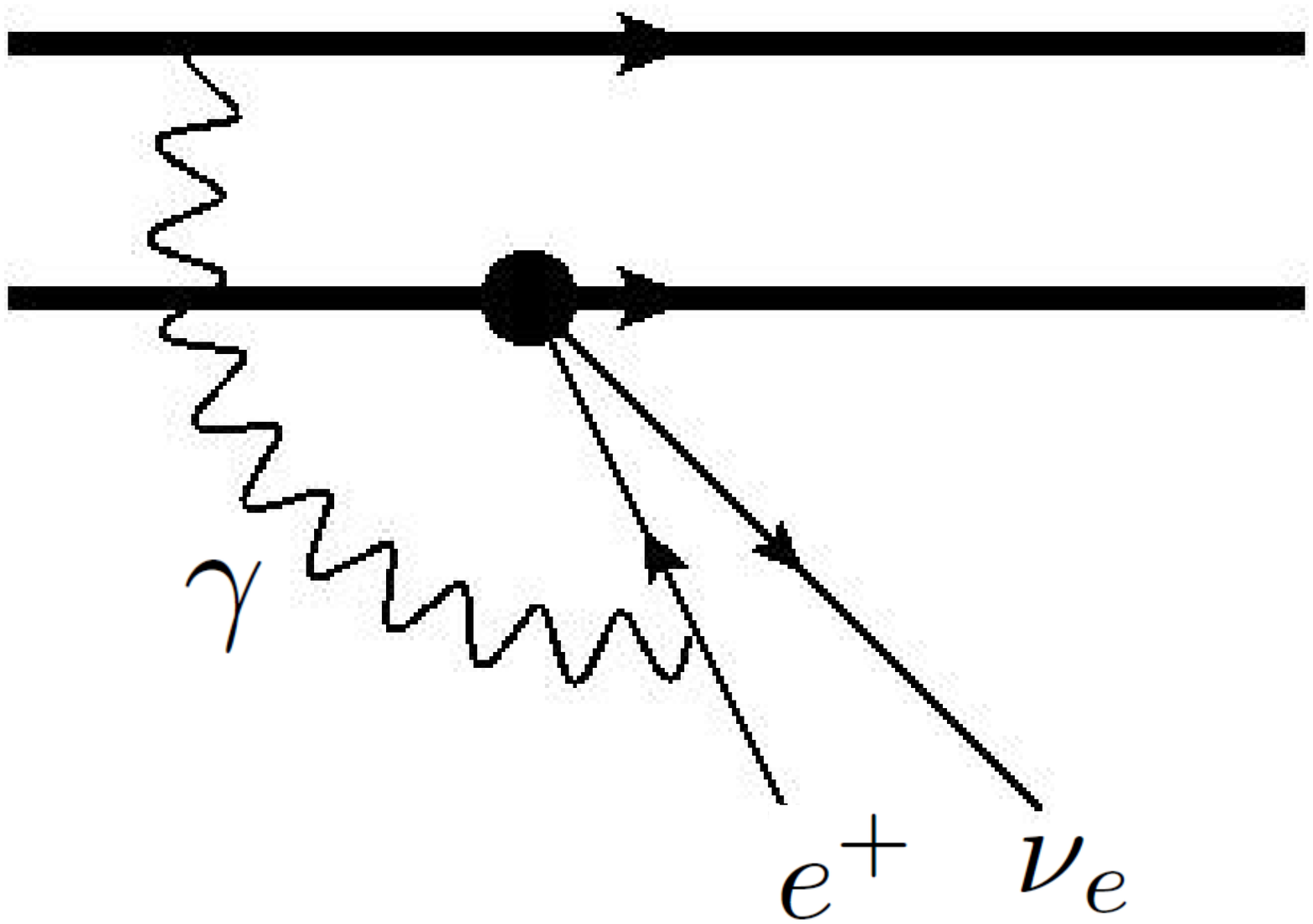}
		\par\end{centering}
	\caption{\label{fig:deltaNS}Representative diagrams for the long-range RC in nuclear $\beta$ decays.}
\end{figure}

Following the seminal work by Sirlin~\cite{Sirlin:1977sv}, the $\mathcal{O}(\alpha)$ electroweak radiative correction (EWRC) in a generic semileptonic $\beta$ decay gives rise to the following multiplicative factor to the decay rate: 
\begin{equation}
\left(1+\frac{\alpha}{2\pi}\bar{g}\right)\left[1+\frac{\alpha}{2\pi}\left(4\ln\frac{M_Z}{m_p}+\ln\frac{m_p}{M_A}+2C+\mathcal{A}_g\right)\right]~,
\end{equation}
where $M_A$ denotes an infrared scale below which non-perturbative Quantum Chromodynamics (QCD) takes place, and $\mathcal{A}_g$ represents a perturbative QCD correction. The quantity $C$ characterizes the long-distance part of the RC originates from the so-called nuclear $\gamma W$-box diagram and is governed by non-perturbative QCD. Historically, one splits it into two terms:
\begin{equation}
C=C_\text{Born}+C_\text{NS}~,
\end{equation}
where $C_\text{Born}$ comes from the first two diagrams in Fig.\ref{fig:deltaNS} which involves only a single active nucleon; $C_\text{NS}$, on the other hand, comes from the third diagram which involves two active nucleons and is an intrinsically nuclear structure effect. 
$C_\text{NS}$ was first introduced in Ref.\cite{Jaus:1989dh} and calculated with shell models in Refs.\cite{Barker:1991tw,Towner:1992xm}.
On the other hand, Ref.\cite{Towner:1994mw}
took into account the experimental fact that the single-nucleon coupling constants for spin-flip processes are quenched in the nuclear medium~\cite{Brown:1983zzc,Brown:1985zz,Brown:1987obh,Towner:1987zz}, and computed $C_\text{Born}$ using the quenched coupling constants:
\begin{equation}
C_\text{Born}=C_\text{Born}^\text{free}+(q_Aq_S^{(0)}-1)C_\text{Born}^\text{free}~,
\end{equation}
where $C_\text{Born}^\text{free}$ is the free nucleon version of $C_\text{Born}$, and $q_A$, $q_S^{(0)}$ are the quenching factor for the axial coupling and the isoscalar spin-magnetic moment coupling, respectively. This idea was subsequently applied to $C_\text{NS}$, which was then recalculated in Ref.\cite{Towner:2002rg,Towner:2007np} using the quenched operators:
$C_\text{NS}\rightarrow C_\text{NS}^\text{quenched}$. The treatments above give:
\begin{equation}
\delta_\text{NS}=\frac{\alpha}{\pi}(q_Aq_S^{(0)}-1)C_\text{Born}^\text{free}+\frac{\alpha}{\pi}C_\text{NS}^\text{quenched}~,
\end{equation}
which is the commonly-adopted representation of $\delta_\text{NS}$ in the global analysis of superallowed $\beta$ decays~\cite{Hardy:2008gy,Hardy:2014qxa}.

\begin{figure}[tb]
	\begin{centering}
		\includegraphics[scale=0.3]{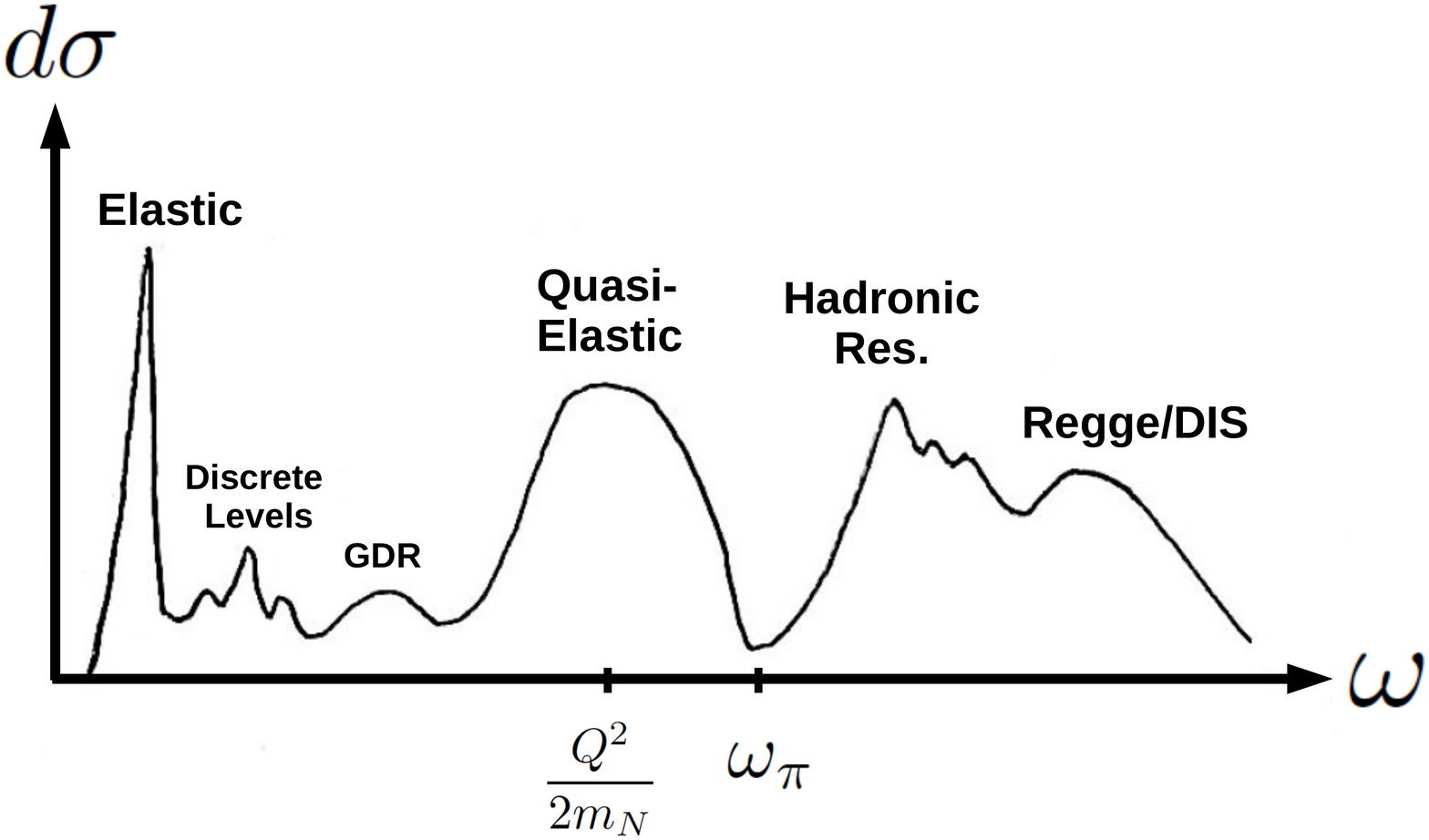}
		\par\end{centering}
	\caption{\label{fig:absorption}A rough sketch of the nuclear absorption spectrum, with $\omega$ the photon energy.}
\end{figure}

\begin{table}
	\begin{centering}
		\begin{tabular}{|c|c|}
			\hline 
			Parent Nucleus & $\delta_{\text{NS}}$(\%)\tabularnewline
			\hline 
			\hline 
			$^{10}$C & -0.400(50)\tabularnewline
			\hline 
			$^{14}$O & -0.283(64)\tabularnewline
			\hline 
			$^{18}$Ne & -0.326(55)\tabularnewline
			\hline 
			$^{22}$Mg & -0.250(50)\tabularnewline
			\hline 
			$^{26}$Si & -0.234(54)\tabularnewline
			\hline 
			$^{30}$S & -0.195(56)\tabularnewline
			\hline 
			$^{34}$Ar & -0.181(60)\tabularnewline
			\hline 
			$^{38}$Ca & -0.167(64)\tabularnewline
			\hline 
			$^{42}$Ti & -0.233(68)\tabularnewline
			\hline 
			$^{46}$Cr & -0.164(72)\tabularnewline
			\hline 
			$^{50}$Fe & -0.140(75)\tabularnewline
			\hline 
			$^{54}$Ni & -0.143(79)\tabularnewline
			\hline 
		\end{tabular} %
		\begin{tabular}{|c|c|}
			\hline 
			Parent Nucleus & $\delta_{\text{NS}}$(\%)\tabularnewline
			\hline 
			\hline 
			$^{26m}$Al & -0.019(51)\tabularnewline
			\hline 
			$^{34}$Cl & -0.093(57)\tabularnewline
			\hline 
			$^{34m}$K & -0.098(60)\tabularnewline
			\hline 
			$^{42}$Sc & 0.033(64)\tabularnewline
			\hline 
			$^{46}$V & -0.031(65)\tabularnewline
			\hline 
			$^{50}$Mn & -0.029(69)\tabularnewline
			\hline 
			$^{54}$Co & -0.017(74)\tabularnewline
			\hline 
			$^{62}$Ga & -0.016(82)\tabularnewline
			\hline 
			$^{66}$As & -0.030(85)\tabularnewline
			\hline 
			$^{70}$Br & -0.049(89)\tabularnewline
			\hline 
			$^{74}$Rb & -0.032(94)\tabularnewline
			\hline 
		\end{tabular}
		\par\end{centering}
	\caption{\label{tab:deltaNS} Current estimates of $\delta_\text{NS}$ in the most recent global analysis of $|V_{ud}|_{0^+}$~\cite{Hardy:2020qwl}.}
	
\end{table}

Recently, a novel dispersion relation (DR) treatment was introduced to study the single-nucleon and nuclear axial $\gamma W$-box diagram~\cite{Seng:2018yzq,Seng:2018qru}. In this method, the box diagram is expressed in terms of an integral over the parity-odd, spin-independent nuclear structure function $F_3(\nu,Q^2)$. At large $Q^2$, it continues smoothly to the single-nucleon structure function so the nuclear and single-nucleon inner RC share the same asymptotic piece.  

The focus, then, is on small and moderate $\{\nu,Q^2\}$ where the two structure functions start to deviate. One may infer the dominant intermediate state contribution to $F_3$ from the general knowledge of the nuclear absorption spectrum, as depicted in Fig.\ref{fig:absorption}. For spinless nuclei, the elastic intermediate state does not contribute to $F_3$ due to parity. Meanwhile, the quenched coupling constants adopted in Ref.\cite{Towner:1994mw} are obtained from nuclear Gamow-Teller transitions and nuclear magnetic moments, which characterize the contributions from the low-lying discrete energy levels at the lower end of Fig.\ref{fig:absorption}. This treatment disregards potentially important contributions from the quasi-elastic absorption peak, as pointed out in Ref.\cite{Seng:2018qru}. 
Another new effect is associated with the electron energy $E_e$.
The $E_e$-dependence of the RC has long been believed negligible, based on the naive dimensional analysis: it can only enter in powers of $E_e/\Lambda$ with $\Lambda$ a relevant strong-interaction scale. For the free nucleon, the lightest choice of such a scale is $M_\pi$, so the energy-dependent corrections scale as $(\alpha/\pi)(E_e/M_\pi)\sim 10^{-5}$, well below the current precision goal. For nuclear systems, as Ref.\cite{Gorchtein:2018fxl} pointed out, the nuclear excitation spectrum features a much lower characteristic scale $\Lambda_\text{nucl}\sim10$\,MeV, comparable to the electron energy available in the decay process. 

Both novel effects were estimated in the free Fermi gas model, demonstrating that they are non-negligible at the precision level of 0.01\%. They tend to partially cancel each other, hardly affecting the central value of $\mathcal{F}t$. However, this cancellation is delicate and model-dependent, and each individual shift is at the level of three standard deviations in terms of the previous analysis of $\delta_\text{NS}$~\cite{Hardy:2008gy,Hardy:2014qxa}.
A conservative analysis resulted in inflated theory uncertainty in $\delta_\text{NS}$~\cite{Gorchtein:2018fxl}, currently the largest in the $|V_{ud}|_{0^+}$ error budget~\cite{Hardy:2020qwl}. A summary of the $\delta_\text{NS}$ values used in global analysis~\cite{Hardy:2020qwl} that includes the aforementioned new nuclear structure uncertainties can be found in Table~\ref{tab:deltaNS}.

In this work we provide the fully-relativistic theory framework to study $\delta_\text{NS}$, based on  the dispersive representation of the nuclear $\gamma W$-box diagram. In this method both the energy-independent and energy-dependent part of $\delta_\text{NS}$ are simultaneously taken into account. To proceed further, one must compute the parity-odd nuclear structure function $F_3(\nu,Q^2)$, which can be obtained from the discontinuity of nuclear matrix elements involving the nuclear Green's function
$G(\omega)=1/(H_0-\omega)$, with $H_0$ the (isospin-symmetric) QCD Hamiltonian. 
A recent proposal to study $\delta_\text{C}$ with ab-initio methods also involves computing the nuclear Green's function~\cite{Seng:2022epj}. 
A program on calculating $\delta_\text{NS}$ and $\delta_\text{C}$ with modern ab-initio methods will enable one to assess the nuclear structure uncertainties in $V_{ud}$ in a controllable way.

%The contents of this paper are arranged as follows. In Section~\ref{sec:notations} we introduce the basic notations for currents, amplitudes and tensors. In Section~\ref{sec:general} we review the general description of RC in superallowed $\beta$ decays based on Sirlin's representation, which allows us to rigorously define $\delta_\text{NS}$. In Section~\ref{sec:crossing} we discuss the $\nu\rightarrow-\nu$ crossing symmetry of the invariant amplitudes in the generalized Compton tensors which is crucial for the evaluation of the loop integrals. In Section~\ref{sec:DR} we introduce the dispersive representation of the single nucleon and nuclear $\gamma W$-box diagram, and in Section~\ref{sec:strategy} we briefly discuss possible strategies to tackle the nuclear box diagram and to deduce $\delta_\text{NS}$, and introduce in Section~\ref{sec:isospin} some useful isospin rotation formula to facilitate this process. In Section~\ref{sec:nuclear} we recast the results above in terms of notations more familiar to the nuclear physics community. We summarize in Section~\ref{sec:final} and provide some technical details in Appendix~\ref{sec:tree}--\ref{sec:MultipoleApp}. 

\section{\label{sec:notations}Basic notation}

The discussion of the EWRC relies heavily on isospin symmetry, so it is useful to introduce vector and axial currents that transform as irreducible tensors in the isospin space, and use them as the basis to construct the physical EW currents. Concentrating on the light quark doublet $\psi=(u,d)^\mathrm{T}$, we define the vector and axial isospin currents as:
\begin{equation}
V_{Im_I}^\mu=\bar{\psi}\gamma^\mu\Gamma_{Im_I}\psi~,~A_{Im_I}^\mu=\bar{\psi}\gamma^\mu\gamma_5\Gamma_{Im_I}\psi~,\label{eq:vectoraxial}
\end{equation}
where $\Gamma_{Im_I}$ are matrices in the isospin space\footnote{Throughout this paper we adopt the particle physics convention of isospin, namely $m_I=+1/2$ for proton and $-1/2$ for neutron. Notice that some nuclear physics papers adopt the opposite convention.}:
\begin{equation}
\Gamma_{00}=\mathbb{I}_2~,~\Gamma_{10}=\tau_3~,~\Gamma_{1\pm 1}=\mp\frac{\tau_1\pm i\tau_2}{\sqrt{2}}~.
\end{equation}
Thus the SM EW currents in the quark sector can be expressed as:
\begin{eqnarray}
J_\mathrm{em}^\mu&=&J_\mathrm{em}^{(0)\mu}+J_\mathrm{em}^{(1)\mu}=\frac{1}{6}V_{00}^\mu+\frac{1}{2}V_{10}^\mu\nonumber\\
J_W^{\mu}&=&(J_W^{\mu})_V+(J_W^{\mu})_A=-\frac{1}{\sqrt{2}}V_{1+1}^\mu+\frac{1}{\sqrt{2}}A_{1+1}^\mu\nonumber\\
J_Z^\mu&=&(J_Z^\mu)_V+(J_Z^\mu)_A=\left(1-2\sin^2\theta_W\right)V_{10}^\mu-\frac{2}{3}\sin^2\theta_W V_{00}^\mu-A_{10}^\mu~,
\end{eqnarray} 
where $\theta_W$ is the weak mixing angle. Following standard notations, we split the electromagnetic (EM) current into isoscalar and isovector components, and the weak currents into vector and axial components. Also, in order to describe $\beta^\pm$ decays simultaneously, we adopt the following notation:
\begin{eqnarray}
J^\mu\equiv (J^\mu)_V+(J^\mu)_A & = & \left\{ \begin{array}{ccc}
J_W^{\dagger\mu}=+\frac{1}{\sqrt{2}}V_{1-1}^\mu-\frac{1}{\sqrt{2}}A_{1-1}^\mu & , & \beta^+-\text{decay}\\
J_W^\mu=-\frac{1}{\sqrt{2}}V_{1+1}^\mu+\frac{1}{\sqrt{2}}A_{1+1}^\mu & , & \beta^--\text{decay}
\end{array}\right.~.
\end{eqnarray}

%\subsection{Tree-level decay amplitude}

The tree-level amplitude of the $\phi_i(p)\rightarrow \phi_f(p')+e^+(p_e)+\nu_e(p_\nu)$ decay process reads,
\begin{equation}
\mathfrak{M}_0=-\frac{G_F}{\sqrt{2}}V_{ud}^*L_\lambda F^\lambda(p',p)~.\label{eq:tree}
\end{equation}
where $L^\lambda=\bar{u}_e\gamma^\lambda(1-\gamma_5)v_\nu$ is the leptonic current. Meanwhile, upon neglecting the spin of the external strongly-interacting particles, the hadronic matrix element of the charged weak current is expressed in terms of two form factors:
\begin{equation}
F^\lambda(p',p)=\langle\phi_f(p')|J^{\lambda}(0)|\phi_i(p)\rangle=f_+(t)(p+p')^\lambda+f_-(t)(p-p')^\lambda~,
\end{equation}
with $t=(p-p')^2$. Here, $|\phi(p)\rangle$ represents a plane-wave state that is normalized relativistically:
\begin{equation}
\langle \phi(p_2)|\phi(p_1)\rangle=(2\pi)^3 2E_\phi(\vec{p}_1)\delta^{(3)}(\vec{p}_1-\vec{p}_2)~.
\end{equation}
In the exact isospin limit, the normalization of the form factors at $t=0$ is given by $f_+(0)=1(\sqrt{2})$ for $I=1/2(1)$ systems, respectively, while $f_-(0)=0$. Also, throughout this paper we work in the rest frame of the parent particle, i.e. $p^\mu=(M_i,\vec{0})$ unless stated otherwise. 

%\subsection{Nuclear tensors}

A key ingredient for the EWRC is the generalized Compton tensor which involves the time-ordered product between the EM current and the charged weak current:
\begin{equation}
T^{\mu\nu}(q;p',p)\equiv\frac{1}{2}\int d^4x e^{iq\cdot x}\langle \phi_f(p')|T\{J_\mathrm{em}^\mu(x)J^{\nu}(0)\}|\phi_i(p)\rangle~.\label{eq:Tmunu}
\end{equation}
We may also define $\Gamma^\mu(q;p',p)$ by replacing $J^\nu$ in $T^{\mu\nu}$ by $\partial\cdot J$. 
The tensor satisfies the EM and charged weak Ward identities:
\begin{eqnarray}
q_\mu T^{\mu\nu}(q;p',p)&=&-\frac{1}{2}i\eta F^\nu(p',p)~,~\text{EM}\nonumber\\
(p'-p+q)_\nu T^{\mu\nu}(q;p',p)+i\Gamma^\mu(q;p',p)&=&-\frac{1}{2}i\eta F^\mu(p',p)~,~\text{Charged Weak}\label{eq:WIs}
\end{eqnarray}
where $\eta=\pm 1$ for $\beta^\pm$-decays. 

We are particularly interested in the forward limit $T^{\mu\nu}(p,q)\equiv T^{\mu\nu}(q;p,p)$ assuming exact isospin symmetry, $M_i=M_f=M$. 
Its discontinuity with respect to the variable $\nu\equiv p\cdot q/M=q_0$ is, at $\nu >0$, given by:
\begin{equation}
\text{Disc}T^{\mu\nu}\equiv T^{\mu\nu}(\nu+i\varepsilon)-T^{\mu\nu}(\nu-i\varepsilon)=4\pi W^{\mu\nu}~,
\end{equation}
where the on-shell tensor $W^{\mu\nu}$ in the forward limit reads:
\begin{eqnarray}
W^{\mu\nu}(p,q)&\equiv&\frac{1}{8\pi}\int d^4xe^{iq\cdot x}\langle \phi_f(p)|[J^\mu_\mathrm{em}(x),J^{\nu}(0)]|\phi_i(p)\rangle\nonumber\\
&=&\frac{1}{8\pi}\sum_X(2\pi)^4\delta^{(4)}(p+q-p_X)\langle\phi_f(p)|J_\text{em}^\mu(0)|X\rangle\langle X|J^{\nu}(0)|\phi_i(p)\rangle\label{eq:Wmunu}~,
\end{eqnarray}
where $X$ runs over all possible intermediate states. 
As a consequence of the splitting $J^{\nu}=(J^{\nu})_V+(J^{\nu})_A$, we may split $T^{\mu\nu}=T^{\mu\mu}_V+T^{\mu\nu}_A$ (and the same for $W^{\mu\nu}$) accordingly.

For the vector component, it is useful to split it into the ``Born'' (B) and ``non-Born'' (nB) pieces: $T^{\mu\nu}_V=T^{\mu\nu}_{V,\text{B}}+T^{\mu\nu}_{V,\text{nB}}$. The Born piece by itself satisfies the full WIs in Eq.\eqref{eq:WIs}~\cite{Seng:2022tjh}, so the non-Born piece satisfies the following homogeneous WIs:
\begin{equation}
q_\mu T^{\mu\nu}_{V,\text{nB}}=(p'-p+q)_\mu T^{\mu\nu}_{V,\text{nB}}=0~.
\end{equation}
Here we have set $\Gamma^V_\mu=0$ by assuming exact isospin symmetry. Therefore, in the forward limit there are only two spin-independent structures:
\begin{eqnarray}
T^{\mu\nu}_{V,\text{nB}}(p,q)&=&\left(-g^{\mu\nu}+\frac{q^\mu q^\nu}{q^2}\right)T_1(\nu,Q^2)+\frac{\hat{p}^\mu\hat{p}^\nu}{M\nu}T_2(\nu,Q^2)\nonumber\\
W^{\mu\nu}_{V,\text{nB}}(p,q)&=&\left(-g^{\mu\nu}+\frac{q^\mu q^\nu}{q^2}\right)F_1(\nu,Q^2)+\frac{\hat{p}^\mu\hat{p}^\nu}{M\nu}F_2(\nu,Q^2)~,\label{eq:invamp12}
\end{eqnarray}
where $\hat{p}^\mu=p^\mu-p\cdot q q^\mu/q^2$, and $T_i$, $F_i$ are Lorentz scalar functions (we call them ``invariant amplitudes'' and ``structure functions'', respectively) of $\nu$ and $Q^2\equiv-q^2$. 

Finally, the axial component is particularly important for the so-called ``inner'' RC to the Fermi matrix element of semileptonic $\beta$ decays. Due to parity, there is only one spin-independent structure for the forward tensors:
\begin{equation}
T^{\mu\nu}_A(p,q)=\frac{i\epsilon^{\mu\nu\alpha\beta}p_\alpha q_\beta}{2M\nu}T_3(\nu,Q^2)~,~W^{\mu\nu}_A(p,q)=\frac{i\epsilon^{\mu\nu\alpha\beta}p_\alpha q_\beta}{2M\nu}F_3(\nu,Q^2)~,\label{eq:invamp}
\end{equation}
where $\epsilon^{0123}=-1$. 

\section{\label{sec:general}General structure of RC in superallowed nuclear decays}

To extract $|V_{ud}|$ from superallowed $\beta$ decays, we make use of the master formulas~\eqref{eq:Vudmaster},\eqref{eq:Ftdef}, 
with a brief derivation provided in Appendix~\ref{sec:tree}. At the precision level of $10^{-4}$, there are four higher-order SM corrections that enter the formula: the nuclear-structure-independent inner RC $\Delta_R^V$, the nuclear-structure-dependent outer and inner RC $\delta_\mathrm{R}'$ and $\delta_\text{NS}$,   respectively, and the ISB correction to the value of $f_+(0)$ $\delta_\text{C}$. 
%Among them, the outer correction $\delta_\text{R}'$ was computed to high accuracy~\cite{Sirlin:1967zza,Sirlin:1987sy,Sirlin:1986cc}, while $\delta_\text{C}$ is beyond the scope of this paper; 
Our discussion in this work regards a simultaneous account of the two inner RC, structure-independent $\Delta_R^V$ and structure-dependent $\delta_\text{NS}$ for the specific case of superallowed $\beta$ decays. 

An appropriate theory framework to study the  EWRC to a generic semileptonic $\beta$ decay is the Sirlin's representation~\cite{Sirlin:1977sv,Seng:2021syx}. Within this framework, the $\mathcal{O}(\alpha)$ virtual correction to the decay amplitude can be cast in the following form,
\begin{eqnarray}
\delta\mathfrak{M}_v&=&\left\{\frac{\alpha}{4\pi}\left[3\ln\frac{M_Z}{m_p}+\ln\frac{M_Z}{M_W}+\tilde{a}_g\right]+\frac{1}{2}\delta_\mathrm{HO}^\mathrm{QED}\right\}\mathfrak{M}_0\nonumber\\
&&+\frac{\alpha}{4\pi}\left\{3\ln\frac{m_p}{m_e}+2\ln\frac{m_e}{M_\gamma}-\frac{3}{4}\right\}\mathfrak{M}_0+\left(\delta\mathfrak{M}_2+\delta\mathfrak{M}_{\gamma W}^a\right)_\mathrm{int}+\delta\mathfrak{M}_3+\delta\mathfrak{M}_{\gamma W}^b~,\label{eq:Sirlin}
\end{eqnarray}
where $M_\gamma$ is a fictitious photon mass to regularize the infrared divergence. Quantities in the expression that depend on non-perturbative QCD are $\delta\mathfrak{M}_{\gamma W}^{a,b}$ which come from the $\gamma W$-box diagram, and $\delta\mathfrak{M}_{2,3}$ which come from the electromagnetic radiative corrections (EMRC) to the nuclear weak current. 
Throughout this work we restrict ourselves to the spin-independent part of $\delta\mathfrak{M}_v$. 
The first line in Eq.\eqref{eq:Sirlin} is universal to all semileptonic $\beta$ decays, while the various terms in the second line are,
\begin{eqnarray}
(\delta\mathfrak{M}_2+\delta \mathfrak{M}_{\gamma W}^a)_\text{int}&=&\sqrt{2}\eta G_F e^2V_{ud}^*L_\lambda\int\frac{d^4q}{(2\pi)^4}\frac{M_W^2}{M_W^2-q^2}\frac{1}{(p_e-q)^2-m_e^2}\frac{1}{q^2-M_\gamma^2}\nonumber\\
&&\times\left\{\frac{2p_e\cdot q q^\lambda}{q^2-M_\gamma^2}T^\mu_{\:\:\mu}+2p_{e\mu}T^{\mu\lambda}-(p-p')_\mu T^{\lambda\mu}+i\Gamma^\lambda\right\}\,,\nonumber\\
\delta \mathfrak{M}_{\gamma W}^{b}&=&-i\sqrt{2}G_F e^2V_{ud}^* L_\lambda\int\frac{d^4q}{(2\pi)^4}\frac{M_W^2}{M_W^2-q^2}\frac{\epsilon^{\mu\nu\alpha\lambda}q_\alpha}{[(p_e-q)^2-m_e^2]q^2}T_{\mu\nu}~.
\end{eqnarray}
The three-point function $\delta\mathfrak{M}_3$ gives a negligible contribution to the Fermi matrix element in the forward limit and assuming isospin symmetry.
The first two bracketed terms of the second line, together with the bremsstrahlung corrections that we do not spell out explicitly, give rise to the well-calculated outer correction $\delta_\mathrm{R}'$ and the $\mathcal{O}(Z_f\alpha)$ piece in the Fermi function, as well as a very small nuclear structure correction which we will see later.

\begin{figure}[tb]
	\begin{centering}
		\includegraphics[scale=0.2]{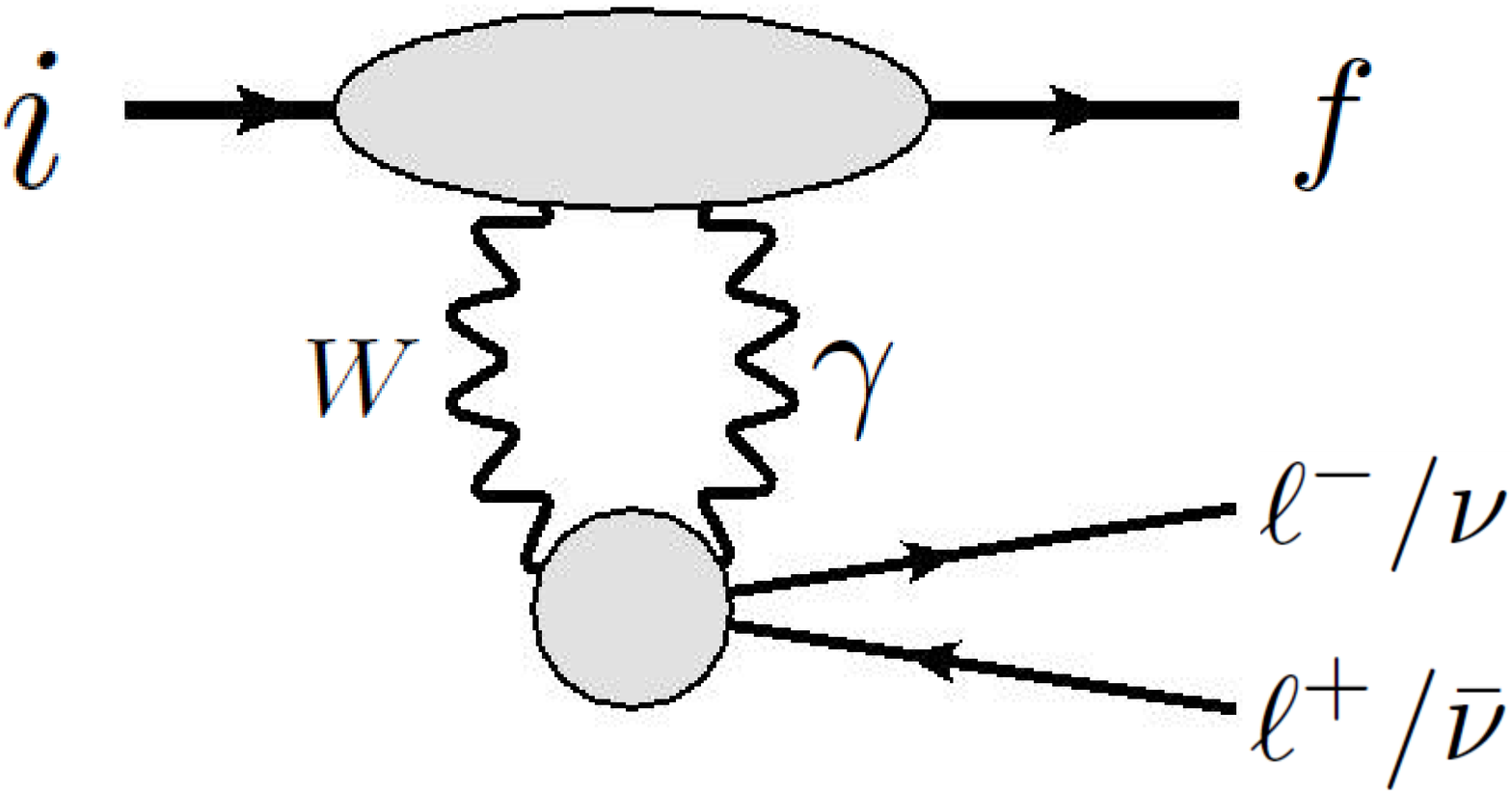}
		\par\end{centering}
	\caption{\label{fig:box}The $\gamma W$-box diagram in $\beta^\pm$ decays.}
\end{figure}

The piece that contains the largest QCD uncertainty is $\delta\mathfrak{M}_{\gamma W}^b$, namely the part of the $\gamma W$-box diagram (see Fig.\ref{fig:box}) that contains an $\epsilon$-tensor coming from the lepton spinor structure.
We may further split this amplitude as $\delta \mathfrak{M}_{\gamma W}^b=\delta \mathfrak{M}_{\gamma W}^{b,V}+\delta \mathfrak{M}_{\gamma W}^{b,A}$ which results from the same splitting in $T^{\mu\nu}$.
Since the spin-independent piece in $T^{\mu\nu}_V$ is symmetric in the Lorentz indices, the contribution of $\delta \mathfrak{M}_{\gamma W}^{b,V}$ must take the form of $\epsilon^{\mu\nu\alpha\lambda} p_\mu p'_\nu p_{e\alpha}L_\lambda$, which vanishes in the forward limit. Hence, the only non-trivial piece for free nucleon and nuclei is $\delta\mathfrak{M}_{\gamma W}^{b,A}$.
To proceed, we make two more approximations in $\delta\mathfrak{M}_{\gamma W}^{b,A}$:
\begin{enumerate}
	\item We make a non-recoil approximation $T_{\mu\nu}^A(q;p',p)\approx T_{\mu\nu}^A(q;p,p)\equiv T_{\mu\nu}^A(p,q)$. This is well-justified in the inner RC for $\beta$ decays of nearly-degenerate systems.
	\item We are allowed to take $m_e\rightarrow 0$ in the electron propagator since $\delta \mathfrak{M}_{\gamma W}^{b,A}$ is free from electron mass singularities~\cite{Seng:2022tjh}. 
\end{enumerate}
With these approximations, $\delta\mathfrak{M}_{\gamma W}^{b,A}$ is proportional to the tree-level amplitude:
\begin{equation}
\delta \mathfrak{M}_{\gamma W}^{b,A}\approx \Box^b_{\gamma W}\mathfrak{M}_0~,\label{eq:gammaWprop}
\end{equation}
and the multiplicative factor is expressed in terms of the parity-odd amplitude $T_3$:
\begin{equation}
\Box^b_{\gamma W}=-e^2\int\frac{d^4q}{(2\pi)^4}\frac{M_W^2}{M_W^2+Q^2}\frac{1}{(p_e-q)^2Q^2}\frac{Q^2+M\nu\frac{p_e\cdot q}{p\cdot p_e}}{M\nu}\frac{T_3(\nu,Q^2)}{f_+(0)}~.\label{eq:BoxgammaW}
\end{equation}

For the $(\delta \mathfrak{M}_2+\delta\mathfrak{M}_{\gamma W}^a)_\text{int}$ term, the outer correction and the Fermi function contain all IR-sensitive contributions that stem from the Born term $T^{\mu\nu}_{V,\text{B}}$. The remaining part stems from $T^{\mu\nu}_{V,\text{nB}}$ and with the same approximations as before, reads,
\begin{equation}
(\delta\mathfrak{M}_2+\delta\mathfrak{M}_{\gamma W}^a)_\text{int}^\text{nB}\approx\Box_{\gamma W}^a\mathfrak{M}_0~,
\end{equation}
where the superscript denotes that only infrared-finite, non-Born parts are retained. Furthermore, 
\begin{eqnarray}
\Box_{\gamma W}^a&=&-\frac{2e^2}{f_+(0)}\eta\int\frac{d^4q}{(2\pi)^4}\frac{1}{(p_e-q)^2}\left\{\frac{-2(p_e\cdot q)^2}{p\cdot p_e (Q^2)^2}T_1(\nu,Q^2)\right. \nonumber\\
&&\left.+\left[\frac{M(p_e\cdot q)^2}{p\cdot p_e \nu (Q^2)^2}-\frac{2p_e\cdot q}{(Q^2)^2}-\frac{p\cdot p_e}{M\nu Q^2}\right]T_2(\nu,Q^2)\right\}~.\label{eq:gammaWa}
\end{eqnarray}
Notice that this piece vanishes when $E_e\rightarrow 0$. 
The two contributions can be put together, \begin{equation}
\Box_{\gamma W}=\Box_{\gamma W}^a+\Box_{\gamma W}^b.\label{eq:totalbox}
\end{equation} 
As we explained in the Introduction, the $E_e$-dependence in Eq.\eqref{eq:BoxgammaW} may be discarded for a free neutron but not for nuclear systems.

The general formalism above applies to both superallowed decays and the free neutron decay. Using Eqs.\eqref{eq:Sirlin}--\eqref{eq:totalbox}, we express the full inner RC to superallowed nuclear $\beta$ decays as
\begin{equation}
\delta\mathfrak{M}_\text{inner}^\text{nucl}=\left\{\frac{\alpha}{4\pi}\left[3\ln\frac{M_Z}{m_p}+\ln\frac{M_Z}{M_W}+\tilde{a}_g\right]+\frac{1}{2}\delta_\text{HO}^\text{QED}+\Box_{\gamma W}^n\right\}\mathfrak{M}_0+\left\{\Box_{\gamma W}^\text{nucl}(E_e)-\Box_{\gamma W}^n\right\}\mathfrak{M}_0~,\label{eq:innernucl}
\end{equation}
where $\Box_{\gamma W}^n$ and $\Box_{\gamma W}^\text{nucl}(E_e)$ refer to Eq.\eqref{eq:totalbox} for the case of neutron and nuclear $\beta$ decay, respectively; the latter has a  potentially non-negligible energy-dependence. The terms in the first curly brackets coincide with the inner correction to the free neutron $\beta$ decay and are independent of nuclear structure,
\begin{equation}
\Delta_R^V=\frac{\alpha}{2\pi}\left[3\ln\frac{M_Z}{m_p}+\ln\frac{M_Z}{M_W}+\tilde{a}_g\right]+\delta_\text{HO}^\text{QED}+2\Box_{\gamma W}^n~,\label{eq:DeltaRV}
\end{equation}
where $\tilde{a}_g\approx -0.083$ is the represents the perturbative QCD correction not coming from the $\gamma W$-box diagram, and $\delta_\text{HO}^\text{QED}\approx 0.0013$ combines the resummation of the $\mathcal{O}(\alpha)$ Quantum Electrodynamics (QED) logarithms and the leading $\mathcal{O}(\alpha^2)$ corrections~\cite{Czarnecki:2004cw}.
The second curly brackets in Eq.\eqref{eq:innernucl} contain the only structure-dependent part in the inner RC, namely the difference between $\Box_{\gamma W}$ on a nucleus and on a free nucleon. Averaging it over the electron energy spectrum (see Eq.\eqref{eq:statfunc}), but dropping the smaller structure-dependent corrections yields,
\begin{equation}
\delta_\mathrm{NS}=\frac{2\int_{m_e}^{E_m}dE_e|\vec{p}_e|E_e(E_e-E_m)^2F(Z_f,E_e)\left[\mathfrak{Re}\Box_{\gamma W}^\mathrm{nucl}(E_e)-\Box_{\gamma W}^n\right]}{\int_{m_e}^{E_m}dE_e|\vec{p}_e|E_e(E_e-E_m)^2F(Z_f,E_e)}~.\label{eq:deltaNS}
\end{equation}
Eqs.\eqref{eq:DeltaRV} and \eqref{eq:deltaNS} rigorously define the structure-independent and structure-dependent part of the inner RC to superallowed $\beta$ decays in terms of the single-nucleon and nuclear $\gamma W$-box diagram or, conversely, the invariant amplitudes $T_i$ for the free nucleon and nuclei.

\section{\label{sec:crossing}Crossing symmetry of the invariant amplitudes}

The first step towards a complete understanding of the inner RC is to study the symmetry of $T_i(\nu,Q^2)$ under crossing $\nu\rightarrow -\nu$. This is most easily appreciated by observing the expression of $\Box_{\gamma W}$ in the limit of zero electron momentum:
\begin{equation}
\Box_{\gamma W}(0)=\frac{e^2}{f_+(0)}\int\frac{d^4q}{(2\pi)^4}\frac{M_W^2}{M_W^2+Q^2}\frac{Q^2+\nu^2}{Q^4}\frac{T_3(\nu,Q^2)}{M\nu}~.\label{eq:BoxgammaW0E}
\end{equation}
One clearly sees that only the component of $T_3$ that is odd under $\nu\rightarrow-\nu$ contributes to the integral in this limit. Once $E_e\neq 0$ the contribution from $T_{1,2}$ and the even component of $T_3$ start to turn on. Therefore, the knowledge of the crossing symmetry of $T_i$ is essential for the separate discussion of the energy-independent and energy-dependent part of the inner RC. Here we will summarize the main conclusions, leaving the detailed derivation to Appendix~\ref{sec:derivecrossing}.

First, if a certain combination of a product of two currents (which we label generically as $A$) leads to a definite crossing symmetry of the invariant amplitudes, then the crossing can always be described by a factor $\xi^A=\pm 1$, where
\begin{equation}
T_1^A(-\nu,Q^2)=-\xi^AT_1^A(\nu,Q^2)~,~T_{2,3}^A(-\nu,Q^2)=\xi^AT_{2,3}^A(\nu,Q^2)~.
\end{equation}
The usual starting point for the crossing symmetry discussion is to write,
\begin{equation}
T_i(\nu,Q^2)=T_i^{(0)}(\nu,Q^2)+T_i^{(1)}(\nu,Q^2)~,
\end{equation}
following the decomposition $J_\text{em}^\mu=J_\mathrm{em}^{(0)\mu}+J_\mathrm{em}^{(1)\mu}$. Assuming isospin symmetry, we find that
\begin{enumerate}
	\item $T_i^{(0)}$ has a definite crossing symmetry $\xi^{(0)}=-1$ that holds regardless of the external states;
	\item $T_i^{(1)}(\nu,Q^2)$, while vanishing identically for the pion due to G-parity, and having
 a definite crossing symmetry $\xi^{(1)}=+1$ for $I=1/2$ systems, in the general case, and in particular for for $I=1$, has mixed crossing symmetry.
 \end{enumerate}
It is convenient to split $T_i^{(1)}$ into two pieces,
$T_i^{(1)}=T_{i,\text{s}}^{(1)}+T_{i,\text{a}}^{(1)}$, 
defined by symmetrizing (s) and anti-symmetrizing (a) the isotriplet indices, 
	\begin{eqnarray}
	T^{\mu\nu}_{(1),\text{s}}(p,q)&\equiv&\frac{1}{8\sqrt{2}}\int d^4x e^{iq\cdot x}\langle \phi_f(p)|T\{V_{10}^\mu(x) (V_{1-1}^\nu(0)-A_{1-1}^\nu(0))\nonumber\\
	&&+V_{1-1}^\mu(x) (V_{10}^\nu(0)-A_{10}^\nu(0))\}|\phi_i(p)\rangle\nonumber\\
	T^{\mu\nu}_{(1),\text{a}}(p,q)&\equiv&\frac{1}{8\sqrt{2}}\int d^4x e^{iq\cdot x}\langle \phi_f(p)|T\{V_{10}^\mu(x) (V_{1-1}^\nu(0)-A_{1-1}^\nu(0))\nonumber\\
	&&-V_{1-1}^\mu(x) (V_{10}^\nu(0)-A_{10}^\nu(0))\}|\phi_i(p)\rangle
	\label{eq:T31pm}
	\end{eqnarray} 
	They have now definite crossing symmetry, $\xi^{(1),\text{s}}=-1$, $\xi^{(1),\text{a}}=+1$, and we can finally combine all the pieces that possess the same crossing behavior $T_i=T_{i,+}+T_{i,-}$, 
\begin{align}
T_{1,+}&=T_1^{(0)}+T_{1,\text{s}}^{(1)}~,~T_{1,-}=T_{1,\text{a}}^{(1)}~,\\
T_{i,+}&=T_{i,\text{a}}^{(1)}~,~T_{i,-}=T_{i}^{(0)}+T_{i,\text{s}}^{(1)}~~(i=2,3)~.\label{eq:T23evenodd}
\end{align}
The subscript $\pm$ indicates that the function is even or odd under $\nu\rightarrow-\nu$.

Eq.\eqref{eq:T23evenodd} indicates that along with the isoscalar EM current also its isovector part contributes to the energy-independent nuclear box diagram in Eq.\eqref{eq:BoxgammaW0E} through $T_{3,\text{s}}^{(1)}$. The latter is intrinsically a many-body effect: Consider, for example, the two currents $J_\mathrm{em}^{(1)\mu}$ and $(J^{\nu})_A$ acting on the same nucleon, then the symmetrized current product in Eq.\eqref{eq:T31pm} gives zero because $\{\tau_-,\tau_3\}=0$; for the same reason it is also obvious that $T_{3,-}^{(1)}$ cannot exist in the asymptotic regime, where the two currents probe a single quark. The same happens for $I=1/2$ nuclei as the latter can be (roughly) viewed as a single active nucleon on top of an $I=0$ inert core. On the other hand, an $I=1$ nucleus can be viewed as two active nucleons on top of an inert core, so $\tau_-$ and $\tau_3$ can act on different nucleons (i.e. the third diagram in Fig.\ref{fig:deltaNS}); in this case they commute, instead of anti-commute, and give a non-zero contribution to $T_{3,-}^{(1)}$. Also, since the latter originates from the $I=2$ combination of the current product, it cannot be described by a single Regge exchange in the $t$-channel (unlike $T_3^{(0)}$ and $T_{3,+}^{(1)}$) as there is no observed $I=2$ meson. Instead, it manifests itself as local counterterms in a low-energy effective field theory. 
The role of $J_\text{em}^{(1)\mu}$ in $\delta_\text{NS}$ was discussed in Ref.\cite{Towner:1992xm}, but was somehow forgotten or at least not explicitly articulated in later works.

\section{\label{sec:DR}Dispersive representation of the nuclear box diagram}

Although Eqs.\eqref{eq:BoxgammaW}, \eqref{eq:gammaWa} (and its simplified version at $E_e=0$, Eq.\eqref{eq:BoxgammaW0E}) may already serve as a starting point, a dispersive representation of $\Box_{\gamma W}(E_e)$ will prove to be very useful. The structure functions $F_i$ defined in Eqs.\eqref{eq:invamp12}, \eqref{eq:invamp} that appear in the dispersive integral may either be inferred from experimental data, or related to nuclear response functions that are standard objects of study with ab-initio methods. 

Ref.\cite{Gorchtein:2018fxl} provided the first DR expression of the energy-dependent nuclear $\gamma W$-box diagram. There are a few aspects which are further improved in this work:
\begin{enumerate}
	\item Ref.\cite{Gorchtein:2018fxl} considered a forward $\bar{\nu}(p_e)+\phi_i(p)\rightarrow e^+(p_e)+\phi_f(p')$ scattering process. The relation to the actual $\beta$ decay $\phi_i(p)\rightarrow \phi_f(p')+ e^+(p_e)+\nu(p_\nu)$ is not immediately straightforward. In this paper we provide a derivation directly in the $\beta$ decay kinematics, with no additional approximation except from the two discussed in Sec.\ref{sec:general}. 
	\item %Ref.\cite{Gorchtein:2018fxl} only provided approximate expressions up to $\mathcal{O}(E_e)$, while here we give the full $E_e$-dependence. We use the dominant contribution involving $F_3$ as an example. 
 Ref.\cite{Gorchtein:2018fxl} only addressed the energy dependence due to the $T_{2,3,+}$ and $T_{1,-}$ components, which are the only ones surviving in the picture of one active nucleon, relevant for the plane-wave Born approximation used in that Ref. Here we provide a complete formalism.
	\item We fix some typos in Ref.\cite{Gorchtein:2018fxl}, e.g. the sign of the (small) $T_{1,2}$ contributions, and the incorrect use of the active nucleon number, instead of $f_+(0)$, as the normalization. 
\end{enumerate}

We outline the derivation of the DR of $\Box_{\gamma W}^b$ first. Since $T_A^{\mu\nu}(p,q)$ has no poles, it must remain regular for $\nu\rightarrow 0$, hence their dispersion representation reads,
\begin{eqnarray}
iT_{3,-}(\nu,Q^2)&=&4\nu\int_{\nu_\text{thr}}^\infty d\nu'\frac{F_{3,-}(\nu',Q^2)}{\nu^{\prime 2}-\nu^2}\nonumber\\
iT_{3,+}(\nu,Q^2)&=&4\nu^2\int_{\nu_\text{thr}}^\infty d\nu'\frac{F_{3,+}(\nu',Q^2)}{\nu'(\nu^{\prime 2}-\nu^2)}~,\label{eq:DRT3}
\end{eqnarray}
where the full structure function $F_3$ is split as $F_3=F_{3,-}+F_{3,+}$, in complete analogy to $T_3$. The lower integration limit is the minimum value of $\nu'$ such that $F_3(\nu',Q^2)$ is non-zero. For instance, for the one-nucleon knockout one has $\nu_\text{thr}=Q^2/(2M)+\epsilon$, with $\epsilon$ the removal energy.
Using this, we may rewrite Eq.\eqref{eq:BoxgammaW} as:
\begin{eqnarray}
	\Box_{\gamma W}^b(E_e)&=&-\frac{4ie^2}{Mf_+(0)}\int\frac{d^4q}{(2\pi)^4}\frac{M_W^2}{M_W^2+Q^2}\frac{Q^2+M\nu\left(\frac{\nu}{M}-\frac{\vec{p}_e\cdot \vec{q}}{ME_e}\right)}{Q^2[Q^2+2E_e\nu-2\vec{p}_e\cdot\vec{q}-i\varepsilon]}\nonumber\\
	&&\times\int_{\nu_\text{thr}}^\infty d\nu'\left[\frac{F_{3,-}(\nu',Q^2)}{\nu^{\prime 2}-\nu^2}+\frac{\nu F_{3,+}(\nu',Q^2)}{\nu'(\nu^{\prime 2}-\nu^2)}\right]~.\label{eq:boxmod}
\end{eqnarray}

\begin{figure}[tb]
	\begin{centering}
		\includegraphics[scale=0.25]{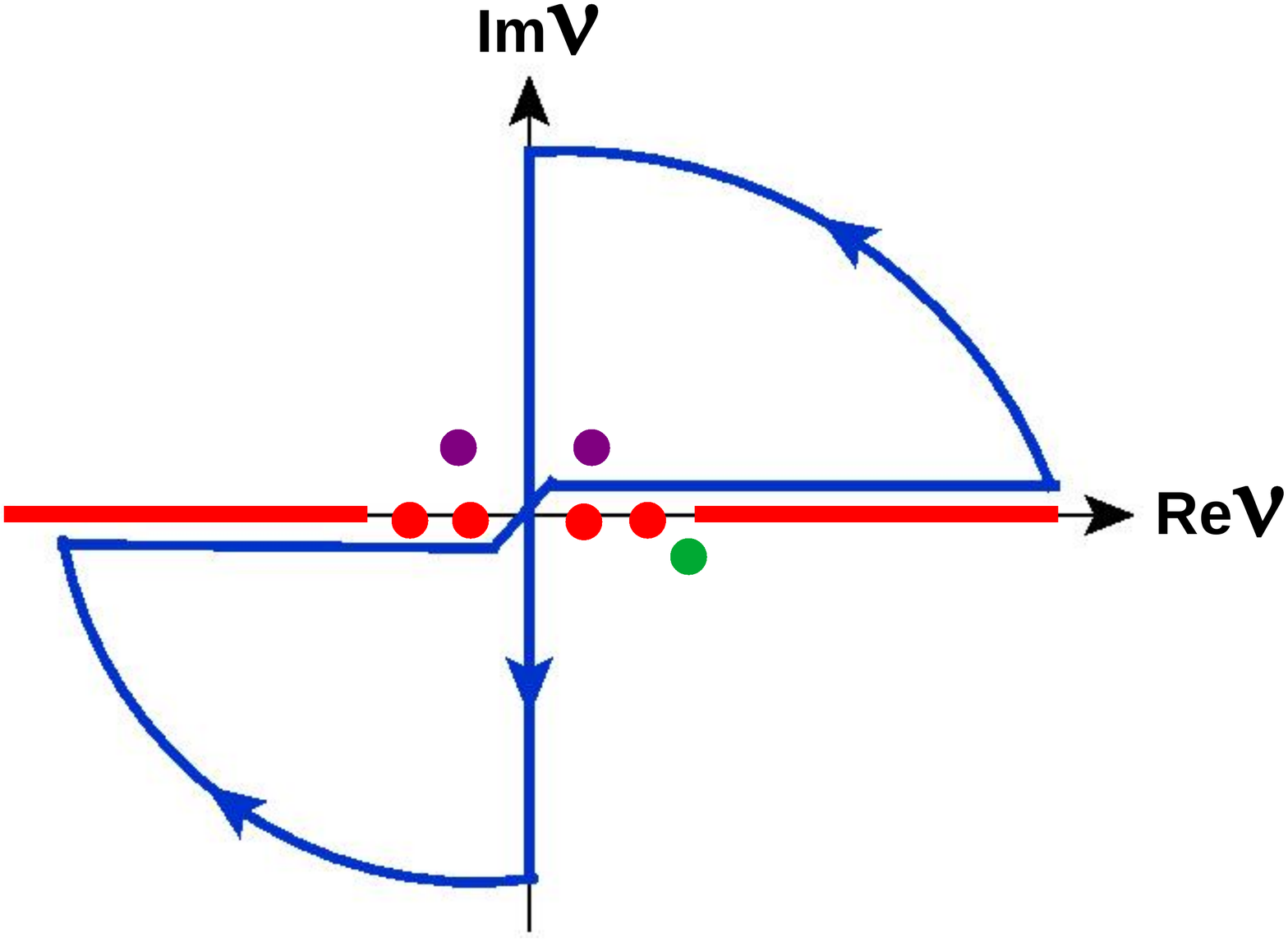}
		\par\end{centering}
	\caption{\label{fig:Wick}(Color online) Blue curve: The Wick rotation contour of the $\nu$-integral. Red lines and dots: Cuts and poles at $\nu=\nu'$. Green dot: The pole $\nu=E_e+|\vec{p}_e-\vec{q}|-i\varepsilon$. Purple dots: Possible positions of the pole $\nu=E_e-|\vec{p}_e-\vec{q}|+i\varepsilon$.  }
\end{figure}

To proceed, we perform the Wick rotation with respect to the variable $\nu$ following the contour depicted in Fig.\ref{fig:Wick}. The electron propagator with non-zero $E_e$ introduces a pole inside the contour, and by Cauchy's theorem
\begin{equation}
\int_{-\infty}^{\infty}d\nu=i\int_{-\infty}^{\infty}d\nu_\mathrm{E}+2\pi i\times\mathrm{Residue}~.
\end{equation}
Therefore, $\Box^b_{\gamma W}(E_e)$ splits into the ``Wick'' contribution and the ``residue'' contribution:
\begin{equation}
\Box^b_{\gamma W}(E_e)=\Box_{\gamma W}^{b,\mathrm{Wick}}(E_e)+\Box_{\gamma W}^{b,\mathrm{res}}(E_e)~,\label{eq:Wickandres}
\end{equation}
and we only need their real part. 

The detailed derivation of $\mathfrak{Re}\Box_{\gamma W}^{b,\text{Wick}}(E_e)$ and $\mathfrak{Re}\Box_{\gamma W}^{b,\text{res}}(E_e)$ can be found in Appendix~\ref{sec:functions}. Denoting the parts that are even and odd functions $E_e$ with the respective superscript and combining the Wick and residue contributions we obtain
\begin{eqnarray}
\mathfrak{Re}\Box_{\gamma W}^{b,\mathrm{even}}(E_e)&=&\frac{\alpha}{2\pi E_e}\frac{1}{Mf_+(0)}\int_0^\infty dQ^2 \frac{M_W^2}{M_W^2+Q^2}\int_{\nu_\text{thr}}^\infty \frac{d\nu'}{\nu'} F_{3,-}(\nu',Q^2)\nonumber\\
&&\times\left\{\ln\left|\frac{E_e+E_\text{min}}{E_e-E_\text{min}}\right|+\frac{\nu'}{2E_e}\ln\left|1-\frac{E_e^2}{E_\text{min}^2}\right|\right\}\nonumber\\
\mathfrak{Re}\Box_{\gamma W}^{b,\mathrm{odd}}(E_e)&=&-\frac{\alpha}{2\pi E_e}\frac{1}{Mf_+(0)}\int_0^\infty dQ^2 \frac{M_W^2}{M_W^2+Q^2}\int_{\nu_\text{thr}}^\infty \frac{d\nu'}{\nu'} F_{3,+}(\nu',Q^2)\nonumber\\
&&\times\left\{\ln\left|1-\frac{E_e^2}{E_\text{min}^2}\right|+\frac{\nu'}{2E_e}\ln\left|\frac{E_e+E_\text{min}}{E_e-E_\text{min}}\right|-\frac{\nu'}{E_\text{min}}\right\}~,\label{eq:DRfull}
\end{eqnarray}
where $E_\text{min}\equiv (\nu'+\sqrt{\nu^{\prime 2}+Q^2})/2$. One finds that the even piece is associated to $F_{3,-}$ and the odd piece to $F_{3,+}$. Finally, a small-$E_e$ expansion gives:
\begin{eqnarray}
\mathfrak{Re}\Box_{\gamma W}^{b,\text{even}}(E_e)&=&\frac{\alpha}{\pi}\int_0^\infty dQ^2\frac{M_W^2}{M_W^2+Q^2}\int_{\nu_\text{thr}}^\infty\frac{d\nu'}{\nu'}\frac{\nu'+2\sqrt{\nu^{\prime 2}+Q^2}}{(\nu'+\sqrt{\nu^{\prime 2}+Q^2})^2}\frac{F_{3,-}(\nu',Q^2)}{Mf_+(0)}+\mathcal{O}(E_e^2)\nonumber\\
\mathfrak{Re}\Box_{\gamma W}^{b,\text{odd}}(E_e)&=&\frac{2\alpha E_e}{3\pi}\int_0^\infty dQ^2\int_{\nu_\text{thr}}^\infty\frac{d\nu'}{\nu'}\frac{\nu'+3\sqrt{\nu^{\prime 2}+Q^2}}{(\nu'+\sqrt{\nu^{\prime 2}+Q^2})^3}\frac{F_{3,+}(\nu',Q^2)}{Mf_+(0)}+\mathcal{O}(E_e^3)\nonumber\\\label{eq:DRapprox}
\end{eqnarray}
which recovers Eq.(10) in Ref.\cite{Gorchtein:2018fxl} upon correcting the typos in the latter. Notice that we removed the factor $M_W^2/(M_W^2+Q^2)$ in $\Box_{\gamma W}^{b,\text{odd}}$ because the integral does not probe the $Q^2\sim M_W^2$ region.

Next we study $\Box_{\gamma W}^a$, with Eq.\eqref{eq:gammaWa} as the starting point. Rather than giving the dispersive representation of $T_{1,\pm}$ and $T_{2,\pm}$ with the full $E_e$-dependence, we retain only the $\mathcal{O}(E_e)$ terms in Eq.\eqref{eq:gammaWa}, with the result
\begin{equation}
\Box_{\gamma W}^a(E_e)=-\frac{4e^2}{3f_+(0)}\frac{E_e}{M}\eta\int\frac{d^4q}{(2\pi)^4}\left\{\frac{4\nu^2+Q^2}{(Q^2)^3}T_{1,+}(\nu,Q^2)+\frac{M(\nu^2+Q^2)}{\nu(Q^2)^3}T_{2,-}(\nu,Q^2)\right\}+\mathcal{O}(E_e^2)~,\label{eq:Boxaapprox}
\end{equation}
where we dropped those components that have the wrong crossing behavior. They satisfy the following DR, 
\begin{align}
T_{1,+}(\nu,Q^2)=\frac{4}{i}\int\limits_{\nu_\text{thr}}^{\infty}\frac{d\nu'\nu'}{\nu^{\prime 2}-\nu^2}F_{1,+}(\nu',Q^2),\quad
T_{2,-}(\nu,Q^2)=\frac{4}{i}\int\limits_{\nu_\text{thr}}^{\infty}\frac{d\nu'\nu}{\nu^{\prime 2}-\nu^2}F_{2,-}(\nu',Q^2)~.
\end{align}
Plugging them into Eq.\eqref{eq:Boxaapprox}, we find,
\begin{eqnarray}
\mathfrak{Re}\Box_{\gamma W}^a(E_e)&=&-\frac{4\alpha}{3\pi }E_e\eta\int_0^\infty dQ^2\int_{\nu_\text{thr}}^{\infty}\frac{d\nu'}{\nu'}\left[\frac{2\nu'}{(\nu'+\sqrt{\nu^{\prime 2}+Q^2})^3}\frac{F_{1,+}(\nu',Q^2)}{ Mf_+(0)}\right.\nonumber\\
&&\left.+\frac{M}{Q^2}\frac{\nu'+2\sqrt{\nu^{\prime 2}+Q^2}}{(\nu'+\sqrt{\nu^{\prime 2}+Q^2})^2}\frac{F_{2,-}(\nu',Q^2)}{ Mf_+(0)}\right]+\mathcal{O}(E_e^2)~.\label{eq:DRapproxa}
\end{eqnarray}
This agrees with the expression in Ref.\cite{Gorchtein:2018fxl} after correcting typos in the latter. 

\section{\label{sec:strategy}Strategy to study $\delta_\text{NS}$}

The dispersive representation provides a useful platform to contrast the free-nucleon and nuclear $\gamma W$-box. To proceed further, we introduce the standard Bjorken variable: $x_B=Q^2/(2M\nu')$, so that we may express the structure functions interchangeably as $F_i(x_B,Q^2)$ or $F_i(\nu',Q^2)$. Next we introduce the Nachtmann moments ($N\geq 1$)~\cite{Nachtmann:1973mr,Nachtmann:1974aj}:
\begin{eqnarray}
M_{i,\pm}(N,Q^2)&\equiv&\frac{N+1}{N+2}\int_0^{x_B^\text{thr}}\frac{dx_B}{x_B^2}\xi^N\left[2x_B-\frac{N\xi}{N+1}\right]\frac{F_{i,\pm}(x_B,Q^2)}{f_+(0)}\nonumber\\
&=&\frac{2}{N+2}\left(\frac{Q^2}{M}\right)^N\int_{\nu_\text{thr}}^{\infty}\frac{d\nu'}{\nu'}\frac{\nu'+(N+1)\sqrt{\nu^{\prime 2}+Q^2}}{(\nu'+\sqrt{\nu^{\prime 2}+Q^2})^{N+1}}\frac{F_{i,\pm}(\nu',Q^2)}{f_+(0)}~,
\end{eqnarray}
where $\xi=2x_B/(1+\sqrt{1+4M^2x_B^2/Q^2})$ and $x_B^\text{thr}=Q^2/(2M\nu_\text{thr})<1$. For asymptotic $Q^2$, the Nachtmann moments reduce to the Mellin moments,
\begin{equation}
\tilde{M}_{i,\pm}(N,Q^2)\equiv \int_0^{1}dx_B x_B^{N-1}\frac{F_{i,\pm}(x_B,Q^2)}{f_+(0)}~,
\end{equation} 
but at finite $Q^2$ they incorporate the target mass corrections. The $\nu'$-integral of $F_2$ and $F_3$ in Eqs.\eqref{eq:DRapprox}, \eqref{eq:DRapproxa} can be recast in the form of Nachtmann moments. For $F_1$ we define a new moment, 
\begin{equation}
M'_{1,+}(2,Q^2)\equiv \left(\frac{Q^2}{M}\right)^2\int_{\nu_\text{thr}}^{\infty}\frac{d\nu'}{\nu'}\frac{2\nu'}{(\nu'+\sqrt{\nu^{\prime 2}+Q^2})^3}\frac{F_{1,+}(x_B,Q^2)}{f_+(0)}~,
\end{equation}
which reduces to the respective Mellin moment at large $Q^2$, $M_{1,+}'(2,Q^2)\rightarrow\tilde{M}_{1,+}(2,Q^2)$. 
In terms of these Nachtmann moments, Eqs.\eqref{eq:DRapprox}, \eqref{eq:DRapproxa} become
\begin{eqnarray}
\mathfrak{Re}\Box_{\gamma W}^{b}(E_e)&=&\frac{3\alpha}{2\pi}\int_0^\infty\frac{dQ^2}{Q^2}\frac{M_W^2}{M_W^2+Q^2}\left[M_{3,-}(1,Q^2)+\frac{8E_e M}{9Q^2}M_{3,+}(2,Q^2)\right]+\mathcal{O}(E_e^2)\nonumber\\
\mathfrak{Re}\Box_{\gamma W}^{a}(E_e)&=&-\frac{4\alpha}{3\pi}\eta\int_0^\infty\frac{dQ^2}{Q^2}\left(\frac{E_e M}{Q^2}\right)\left[M_{1,+}'(2,Q^2)+\frac{3}{2}M_{2,-}(1,Q^2)\right]+\mathcal{O}(E_e^2)~.\label{eq:boxexpand}
\end{eqnarray} 
Assuming these two pieces together give a precise enough description of the nuclear $\gamma W$-box diagram (which needs to be checked by studying its convergence speed), we write,
\begin{eqnarray}
\mathfrak{Re}\Box_{\gamma W}^\text{nucl}-\Box_{\gamma W}^n&\approx&\frac{3\alpha}{2\pi}\int_0^\infty\frac{dQ^2}{Q^2}\Bigl\{\left[M_{3,-}^\text{nucl}(1,Q^2)-M_{3,-}^n(1,Q^2)\right]\nonumber\\
&&+\frac{8E_eM}{9Q^2}\left[M_{3,+}^\text{nucl}(2,Q^2)-\eta M_{1,+}^{\prime\text{nucl}}(2,Q^2)-\frac{3}{2}\eta M_{2,-}^\text{nucl}(1,Q^2)\right]\Bigr\}~.\label{eq:master}
\end{eqnarray}
Above, the factor $M_W^2/(M_W^2+Q^2)$ was removed because the physics at $Q^2\sim M_W^2$ is not probed. As is well-known, the asymptotic contribution to $\Box_{\gamma W}$ is process-independent and cancels between $M_{3,-}^\text{nucl}$ and $M_{3,-}^n$. Plugging this into Eq.\eqref{eq:deltaNS} gives us a closed expression for $\delta_\text{NS}$. Below we discuss some aspects important for evaluating it.\\

%\subsection{Relevant region of $Q^2$}
\noindent
{\bf Relevant region of the $Q^2$-integral}:\\
%\begin{figure}[tb]
%	\begin{centering}
%		\includegraphics[scale=0.4]{nNachtmann}
%		\par\end{centering}
%	\caption{\label{fig:nNachtmann} The free-nucleon first Nachtmann moment $M_{3,-}^n(1,Q^2)$ obtained by combining the elastic contribution with (a) indirect lattice QCD inputs~\cite{Feng:2020zdc} at $Q^2<2~\text{GeV}^2$ (blue band) and (b) four-loop perturbative Quantum Chromodynamics (pQCD) prediction at $Q^2>2~\text{GeV}^2$ (red line)~\cite{Baikov:2010iw,Baikov:2010je}. Details can be found in Ref.\cite{Seng:2020wjq}.  }
%\end{figure}
%
While the integral in Eq.\eqref{eq:master} is insensitive to asymptotically high $Q^2$, we need to find out, starting from which value of $Q^2=Q^2_\text{nucl}$ the cancellation between the nuclear and nucleon boxes is at such a level that a precise enough determination of $\delta_\text{NS}$ can already be obtained with $Q^2_\text{nucl}$ as an upper limit.
The first Nachtmann moment for a free nucleon, $M_{3,-}^n(1,Q^2)$, has been studied recently as a function of $Q^2$ using  phenomenological~\cite{Seng:2018yzq,Seng:2018qru,Hayen:2020cxh,Shiells:2020fqp} and indirect lattice inputs~\cite{Seng:2020wjq}.
%see Fig.\ref{fig:nNachtmann} for a recent example. 
It was found that by $Q^2\approx2$\,GeV$^2$ the perturbative description sets in, and we can expect that 
$Q^2_\text{nucl}<2$\,GeV$^2$. 
A trial calculation of $M_{3,+}^\text{nucl}(1,Q^2)$ at $Q\sim 100-300$~MeV may already provide a useful hint. As evidenced by the entries in Table~\ref{tab:deltaNS}, even a $\sim 10$\% determination of $M_{3,+}^\text{nucl}(1,Q^2)$ will significantly improve the precision of $\delta_\text{NS}$ for most nuclei. \\

%\addtocontents{toc}{\protect\setcounter{tocdepth}{1}}
%\subsection{The energy-dependent corrections}
%\addtocontents{toc}{\protect\setcounter{tocdepth}{2}}
\noindent
{\bf The energy-dependent corrections}:\\
The second line in Eq.\eqref{eq:master} describes the leading $E_e$-dependence in $\delta_\text{NS}$ associated with the structure functions $F_{3,+}^\text{nucl}$, $F_{1,+}^\text{nucl}$ and $F_{2,-}^\text{nucl}$. The parity-even structure functions are analogous to $F_{1,2}^\text{em}$ in ordinary Compton scattering that involve only the EM current and satisfy the Baldin sum rule~\cite{BALDIN1960310}, which relate their moments to the nuclear electric dipole polarizability $\alpha_E$. With a typical estimation of the latter~\cite{Berman:1975tt}, Ref.\cite{Gorchtein:2018fxl} deduced that the contribution of $F_{1,2}$ to $\delta_\text{NS}$ is of the order $10^{-5}$ and could be neglected given our precision goal. One arrives at the same conclusion in the free Fermi gas model: the contribution from $F_3$ is found to be an order of magnitude larger than that from $F_{1,2}$, in part due to the large nucleon isovector magnetic moment. Therefore, from now on we will focus on the parity-odd structure functions $F_{3,\pm}$. \\

%\addtocontents{toc}{\protect\setcounter{tocdepth}{1}}
%\subsection{Possible theoretical approaches}
%\addtocontents{toc}{\protect\setcounter{tocdepth}{2}}
\noindent
{\bf Possible theoretical approaches}:\\
The primary objects of interest are now the nuclear Nachtmann moments $M_3^\text{nucl}(N,Q^2)$ which can be studied in different ways. The more straightforward approach is to compute $F_{3,\pm}(x_B,Q^2)$ directly with ab-initio methods, from which the moments can be evaluated. One of the main challenges is to deal with the sum over all intermediate states in Eq.\eqref{eq:Wmunu}. This can either be done directly, e.g. using the short-time approximation for light nuclei~\cite{Pastore:2019urn}, or indirectly such as using the Lorentz integral transform method~\cite{Efros:1994iq}.

Alternatively, one could compute the amplitudes $T_{3,\pm}$, rather than their discontinuities $F_{3,\pm}$ for $\nu>\nu_\text{thr}$. At low energy $|\nu|<\nu_\text{thr}$ the invariant amplitudes permit a generic low-energy expansion of the form
\begin{eqnarray}
iT_{3,-}(\nu,Q^2)&=&4\sum_{N=0}^\infty\left(\frac{2M\nu}{Q^2}\right)^{2N+1}\tilde{M}_{3,-}(2N+1,Q^2)~~,~~|\nu|<\nu_\text{thr}\nonumber\\
iT_{3,+}(\nu,Q^2)&=&4\sum_{N=0}^\infty\left(\frac{2M\nu}{Q^2}\right)^{2N+2}\tilde{M}_{3,+}(2N+2,Q^2)~~,~~|\nu|<\nu_\text{thr}~,
\end{eqnarray}
with the ``polarizabilities'' expressed as $Q^2$-weighted Mellin moments of $F_3$. Note that negative powers of $Q^2$ are cancelled by higher powers of $Q^2$ implicitly contained in the Mellin moments. 
Ref.\cite{Gorchtein:2021fce} demonstrates, for the case of a single nucleon, how one could accurately reconstruct the desired Nachtmann moment using a few lowest Mellin moments. It is worthwhile exploring the convergence pattern for the case of nuclei.

\section{\label{sec:isospin}Isospin rotation}

All the discussions above are formulated in terms of off-diagonal nuclear matrix elements of products of isospin currents. This may cause inconveniences in practical calculations:
\begin{enumerate}
	\item Although we have assumed exact isospin symmetry of the system, in practice ab-initio methods usually have built-in ISB effects. This may lead to mismatch between the parameters of the initial and final states, e.g. nuclear masses and excitation energies. This could be avoided if the external states were diagonal.
	\item Many existing programs of ab-initio methods were developed to study  observables in electromagnetic or weak neutral current processes that involve physical, instead of isospin, nuclear current operators. 
\end{enumerate}

To trade off-diagonal nuclear matrix elements of isospin currents for diagonal nuclear matrix elements of physical EW currents we invoke isospin rotation by making use of the Wigner-Eckart theorem (WET) in the isospin space (see Appendix~\ref{sec:SU2}); in this section we provide the relevant formulas. 

First, the isospin rotation formula involving the isoscalar vector current is:
\begin{eqnarray}
\langle 1,0|V_{00}^{\mu}\otimes A_{1-1}^\nu|1,1\rangle&=&\langle 1,-1|V_{00}^\mu\otimes A_{1-1}^\nu|1,0\rangle\nonumber\\
&=&-3\left[\langle 1,1|J_\mathrm{em}^{\mu}\otimes(J_Z^{\nu})_A|1,1\rangle-\langle 1,-1|J_\mathrm{em}^{\mu}\otimes(J_Z^{\nu})_A|1,-1\rangle\right]~.\label{eq:isospin0}
\end{eqnarray}
Meanwhile, the formula involving the isovector vector current are:
\begin{eqnarray}
\langle 1,-1|V_{10}^\mu\otimes A_{1-1}^\nu|1,0\rangle &=&2\left[\langle 1,0|(J_W^{\mu})_V\otimes(J_W^{\dagger\nu})_A|1,0\rangle-\langle 1,-1|(J_W^{\mu})_V\otimes(J_W^{\dagger\nu})_A|1,-1\rangle\right]\nonumber\\
\langle 1,0|V_{10}^\mu \otimes A_{1-1}^\nu|1,1\rangle &=&2\left[\langle 1,1|(J_W^{\mu})_V\otimes(J_W^{\dagger\nu})_A|1,1\rangle-\langle 1,0|(J_W^{\mu})_V\otimes(J_W^{\dagger\nu})_A|1,0\rangle\right]
\end{eqnarray}
and 
\begin{eqnarray}
\langle 1,-1|V_{1-1}^\mu\otimes A_{10}^\nu|1,0\rangle &=&2\left[\langle 1,0|(J_W^{\dagger\mu})_V\otimes(J_W^{\nu})_A|1,0\rangle-\langle 1,-1|(J_W^{\dagger\mu})_V\otimes(J_W^{\nu})_A|1,-1\rangle\right]\nonumber\\
\langle 1,0|V_{1-1}^\mu\otimes A_{10}^\nu|1,1\rangle &=&2\left[\langle 1,1|(J_W^{\dagger\mu})_V\otimes(J_W^{\nu})_A|1,1\rangle-\langle 1,0|(J_W^{\dagger\mu})_V\otimes(J_W^{\nu})_A|1,0\rangle\right]
\end{eqnarray}
from which the combinations $T_{3,\pm}^{(1)}$ and $F_{3,\pm}^{(1)}$ can be constructed.

Isospin rotation brings not only practical but also conceptual benefits. 
To illustrate, here we will derive the well-known asymptotic (i.e. large-$Q^2$) contribution to the energy-independent axial box diagram. With its original definition, it is easy to show  using the operator-product-expansion of $T\{J_\text{em}^\mu(x)(J_W^{\dagger\nu}(0))_A\}$ that the asymptotic contribution gives a process-independent large EW logarithm~\cite{Sirlin:1981ie}:
\begin{equation}
\Box_{\gamma W}(0)=\frac{\alpha}{8\pi}\ln\frac{M_W^2}{\Lambda^2}+...\label{eq:largelog}
\end{equation}
where $\Lambda$ is a ultraviolet scale, above which QCD becomes asymptotically-free.
Here we offer an alternative derivation using the dispersive representation. We start from:
\begin{eqnarray}
\mathfrak{Re}\Box_{\gamma W}(0)&=&\frac{\alpha}{\pi}\int_0^\infty dQ^2\frac{M_W^2}{M_W^2+Q^2}\int_{\nu_\text{thr}}^\infty\frac{d\nu'}{\nu'}\frac{\nu'+2\sqrt{\nu^{\prime 2}+Q^2}}{(\nu'+\sqrt{\nu^{\prime 2}+Q^2})^2}\frac{F_{3,-}(\nu',Q^2)}{Mf_+(0)}\nonumber\\
&=&\frac{3\alpha}{2\pi}\int_{\Lambda^2}^{\infty}\frac{dQ^2}{Q^2}\frac{M_W^2}{M_W^2+Q^2}\frac{1}{f_+(0)}\int_0^1dx_B F_3^{(0)}(x_B)+...~.\label{eq:Boxasym}
\end{eqnarray}
In the $Q^2>\Lambda^2$ region the first Nachtmann moment reduces to the first Mellin moment, and only the ``(0)'' component in $F_{3,-}$ survives as we
discussed in Sec.\ref{sec:crossing}. 

We wish to evaluate the $x_B$-integral using parton picture, but the fact that $F_3^{(0)}$ is off-diagonal makes this step less straightforward, so now isospin rotation becomes very useful. For superallowed decays of $I=1$ systems, we use the isospin relation in Eq.\eqref{eq:isospin0} to get:
\begin{equation}
F_3^{(0)}=\frac{1}{2\sqrt{2}}\left[F_{3,+1}^{\gamma Z}-F_{3,-1}^{\gamma Z}\right]~.
\end{equation}
Here $F_{3,\pm 1}^{\gamma Z}$ are defined exactly like $F_3$ in Eq.\eqref{eq:invamp}, except that we replace $(J^{\nu})_A$ in $W_A^{\mu\nu}$ by $(J_Z^\nu)_A$, and the off-diagonal external states by diagonal states with $\{I,m_I\}=\{1,\pm1\}$ respectively.
At large $Q^2$, they could be expressed in terms of the quark distribution functions (see Ref.\cite{Zyla:2020zbs}, but beware of the difference in overall normalization):
\begin{equation}
F_3^{\gamma Z}(x_B)=\frac{1}{3}\left(u(x_B)-\bar{u}(x_B)\right)+\frac{1}{6}\left(d(x_B)-\bar{d}(x_B)\right).
\end{equation}
A nucleus with $Z$ protons and $N$ neutrons thus satisfy:
\begin{equation}
\int_0^1dx_BF_3^{\gamma Z}(x_B)=\frac{1}{3}(2Z+N)+\frac{1}{6}(Z+2N)=\frac{5}{6}Z+\frac{2}{3}N~.
\end{equation}
Recall that for the $m_I=\pm 1$ state we have $Z=A/2\pm 1$ and $N=A/2\mp 1$. Therefore:
\begin{equation}
\int_0^1dx_BF_3^{(0)}(x_B)=\frac{1}{2\sqrt{2}}\int_0^1dx_B\left[F_{3,+1}^{\gamma Z}(x_B)-F_{3,-1}^{\gamma Z}(x_B)\right]=\frac{\sqrt{2}}{12}~.\label{eq:F30sum}
\end{equation}
Finally, we recall that $f_+(0)=\sqrt{2}$ for $I=1$ isomultiplets, so plugging Eq.\eqref{eq:F30sum} into Eq.\eqref{eq:Boxasym} and performing the $Q^2$-integral reproduces the large EW logarithm in Eq.\eqref{eq:largelog}.

The derivation above also highlights a potential caveat in applying the isospin relations: for $A\gg 1$, Eq.\eqref{eq:F30sum} involves a subtraction between two large numbers $\sim A$ to get a number of order 1. In practical nuclear calculations, each of these two large numbers contains some ISB effects which, upon subtraction, may have an unnaturally large effect. We expect this effect to be less problematic for light nuclei.

\section{\label{sec:nuclear}Employing nuclear physics notations}
	
The entire theory framework of box diagram amplitudes and the DR formalism outlined above are inherited from previous studies of single-nucleon EWRC, and thus are based on particle physics notations. Here we re-express them in terms of notations that are more familiar to the nuclear physics community. 

In this section we introduce a somewhat more general notation (we did not do it at the beginning to avoid losing focus on the actual physical problem, i.e. $\beta^+$-decay, that we are interested in, when we discussed the basic framework). We take $J_a^\mu$ as an arbitrary vector current and $J_b^\mu$ as an arbitrary axial current, and define:
\begin{eqnarray}
T^{\mu\nu}_{ab}(p,q)&=&\frac{1}{2}\int d^4x e^{iq\cdot x}\langle \phi_f(p)|T\{J_a^\mu(x)J_b^\nu(0)\}|\phi_i(p)\rangle=\frac{i\epsilon^{\mu\nu\alpha\beta}p_\alpha q_\beta}{2M\nu}T_3^{ab}(\nu,Q^2)\nonumber\\
W^{\mu\nu}_{ab}(p,q)&\equiv&\frac{1}{8\pi}\int d^4xe^{iq\cdot x}\langle \phi_f(p)|[J_a^\mu(x),J_b^\nu(0)]|\phi_i(p)=\frac{i\epsilon^{\mu\nu\alpha\beta}p_\alpha q_\beta}{2M\nu}F_3^{ab}(\nu,Q^2)~,
\end{eqnarray} 
where $\phi_{i,f}$, are spinless nuclear states (diagonal or off-diagonal) with degenerate mass $M$. Taking $J_a^\mu\rightarrow J_\text{em}^\mu$ and $J_b^\mu\rightarrow (J_W^{\dagger\mu})_A$ recovers the $T_A^{\mu\nu}$ and $W_A^{\mu\nu}$ relevant to $\beta^+$-decay, but we may also replace them by isospin currents or other physical EW currents to implement our previous discussions of crossing symmetry and isospin rotation formula. 

\subsection{Connection to nuclear Green's function and nuclear response function}

We will work in the target's rest frame, i.e. $p_0=M$ and $\nu=q_0$, and align $\vec{q}=\q \hat{e}_z$. Using the definition of the time-ordered product,
$T[A(t)B(0)]=\Theta(t)A(t)B(0)+\Theta(-t)B(0)A(t)$, and translational symmetry, 
$A(x)=e^{iP\cdot x}A(0)e^{-iP\cdot x}$,
we can insert a complete set of states between the two currents in the definition of $T_{ab}^{\mu\nu}$, and perform the $t$-integration explicitly, with the result
\begin{eqnarray}
T^{\mu\nu}_{ab}(p,q)&=&-\frac{i}{2}\int d^3x e^{-i\vec{q}\cdot\vec{x}}\langle \phi_f(p)|J_a^\mu(0,\vec{x})G(M+q_0+i\varepsilon)J_b^{\nu}(0,\vec{0})|\phi_i(p)\rangle\nonumber\\
&&-\frac{i}{2}\int d^3x e^{-i\vec{q}\cdot\vec{x}}\langle \phi_f(p)|J_b^{\nu}(0,\vec{0})G(M-q_0+i\varepsilon)J_a^\mu(0,\vec{x})|\phi_i(p)\rangle~.\label{eq:TmunuRel}
\end{eqnarray}
Above, we introduced the nuclear Green's function,
\begin{equation}
G(\omega)\equiv (H_0-\omega)^{-1}=\sum_X\frac{|X\rangle\langle X|}{E_X-\omega}~,
\end{equation}
with $H_0$ the nuclear Hamiltonian. 

The external states in Eq.\eqref{eq:TmunuRel} are relativistically-normalized plane waves, and now we shall rewrite them in terms of properly-normalized states in quantum mechanics. Applying the translational symmetry once again leads to
\begin{eqnarray}
T^{\mu\nu}_{ab}(p,q)&=&-\frac{i}{2V}\langle\phi_f(p)|J_a^\mu(\vec{q})G(M+q_0+i\varepsilon)J_b^{\nu}(-\vec{q})|\phi_i(p)\rangle\nonumber\\
&&-\frac{i}{2V}\langle\phi_f(p)|J_b^{\nu}(-\vec{q})G(M-q_0+i\varepsilon)J_a^\mu(\vec{q})|\phi_i(p)\rangle~,
\end{eqnarray}
where we have defined the (spatial) Fourier transform of currents as
\begin{equation}
J^\mu(\vec{q})\equiv \int d^3x e^{-i\vec{q}\cdot \vec{x}}J^\mu(0,\vec{x})~,
\end{equation}
and the volume factor
\begin{equation}
V\equiv \int d^3x\cdot 1=(2\pi)^3\delta^{(3)}(\vec{0})~,
\end{equation}
with $\vec{0}$ a zero vector in the momentum space. 
Recalling that the plane wave states at $\vec{p}=0$ are normalized as
\begin{equation}
\langle\phi(\vec{0})|\phi(\vec{0})\rangle=2M(2\pi)^3\delta^{(3)}(\vec{0})=2MV~,
\end{equation}
we define a quantum mechanical state
\begin{equation}
|\phi\rangle \equiv \frac{1}{\sqrt{2MV}}|\phi(\vec{0})\rangle
\end{equation}
that normalizes as $\langle \phi|\phi\rangle =1$. This allows us to write
\begin{eqnarray}
T^{\mu\nu}_{ab}(p,q)&=&-iM\langle\phi_f|J_a^\mu(\vec{q})G(M+q_0+i\varepsilon)J_b^{\nu}(-\vec{q})|\phi_i\rangle\nonumber\\
&&-iM\langle\phi_f|J_b^{\nu}(-\vec{q})G(M-q_0+i\varepsilon)J_a^\mu(\vec{q})|\phi_i\rangle~.\label{eq:Tmultipole}
\end{eqnarray}
One sees that the volume factor $V$ drops out, and everything on the right-hand side is rigorously defined in quantum mechanics and is calculable in nuclear theory.
Similarly, one may express the hadronic tensor $W^{\mu\nu}_{ab}$ in the way more familiar in nuclear calculations,
\begin{equation}
W^{\mu\nu}_{ab}(p,q)=\frac{M}{2}\sum_X \delta(q_0+M-E_X)\langle\phi_f|J_a^\mu(\vec{q})|X\rangle\langle X|J_b^{\nu}(-\vec{q})|\phi_i\rangle~,\label{eq:Wmultipole}
\end{equation}  
and the parity-odd structure function reads:
\begin{equation}
F_3^{ab}(\nu,Q^2)=-\frac{i q_0}{\q}\left(W_{ab}^{12}(p,q)-W_{ab}^{21}(p,q)\right)~,\label{eq:F3response}
\end{equation}
in complete analogy to the response function $R_{T'}$ in neutrino-nucleus scattering~\cite{Shen:2012xz,Lovato:2014eva,Lovato:2017cux,Rocco:2020jlx}.

A comment is in order. In studying the invariant amplitudes we encounter nuclear matrix elements of the form $\langle \phi_f|\hat{O}_1 G(\omega)\hat{O}_2|\phi_i\rangle$, where $\hat{O}_{1,2}$ are arbitrary operators. This is a classic problem in nuclear many-body calculations, and is challenging due to the difficulty to invert a large matrix $H_0-\omega$; fortunately, there are ways to circumvent the problem, e.g. using the Lanczos algorithm~\cite{Lanczos:1950zz,Haydock_1974,Marchisio:2002jx}. Interestingly enough, the same form of nuclear matrix elements also appear in a newly-proposed strategy to compute the ISB correction $\delta_\text{C}$~\cite{Seng:2022epj}. 
%In this formalism, one computes a ``generating function'' of the form $\langle \phi|(M_{-1}^{(1)})^\dagger G(\omega) M_{-1}^{(1)}|\phi\rangle$, where $\vec{M}^{(1)}\equiv \sum_i r_i^2 \vec{\tau}(i)/2$ is the so-called isovector monopole operator, with $i$ labeling the nucleons in the nucleus. It was shown that the generating functional itself is related to some ISB combination of electroweak nuclear radii which could be measured, while its $\omega$-derivative is related to $\delta_\text{C}$, so the new formalism provides a way to directly constrain $\delta_\text{C}$ from experiments. 
The fact that $\delta_\text{NS}$ and $\delta_\text{C}$ share the similar form nuclear matrix elements indicates that they could be studied simultaneously given a specific type of an ab-initio method.

\subsection{Multipole expansion of invariant amplitudes and structure functions}

The multipole expansion of nuclear EW currents is a powerful technique frequently adopted in ab-initio studies~\cite{Gazda:2016mrp,Acharya:2019fij,Glick-Magid:2021uwb}. Combining with WET, it converts the problem of calculating the full current matrix elements into the calculation of reduced matrix elements which is more tractable. Here we shall implement the formalism to both the invariant amplitude and structure function for the convenience of interested readers; full details are available in Appendix~\ref{sec:MultipoleApp}.

We start with the invariant amplitude $T_3^{ab}$. Analogous to Eq.\eqref{eq:F3response}, we cast it in an explicitly antisymmetric form:
\begin{equation}
T_3^{ab}(\nu,Q^2)=-\frac{iq_0}{\q}\left(T_{ab}^{12}(p,q)-T_{ab}^{21}(p,q)\right)~.
\end{equation}
Since all currents are transverse (along $\hat{e}_x$ and $\hat{e}_y$ directions),
we may express them in terms of the $\lambda=\pm 1$ components defined in Eq.\eqref{eq:Jqspan}:
\begin{eqnarray}
T_3^{ab}(\nu,Q^2)&=&\frac{iq_0}{\q}M\langle \phi_f|\bigl\{J_a(\vec{q},+1)G(M+q_0+i\varepsilon)J_b(-\vec{q},-1)+\nonumber\\
&&J_b(-\vec{q},-1)G(M-q_0+i\varepsilon)J_a(\vec{q},+1)-(\lambda=+1\leftrightarrow\lambda=-1)\bigr\}|\phi_i\rangle~.\label{eq:T3abint}
\end{eqnarray}
The multipole expansion of both currents reads
\begin{eqnarray}
J_{a/b}(\vec{q},\pm 1)&=&-\sqrt{2\pi}\sum_{J=1}^\infty (-i)^J\sqrt{2J+1}\left(\pm T_{J\pm 1}^{a/b,\mathrm{mag}}(\q)-T_{J\pm 1}^{a/b,\mathrm{el}}(\q)\right)\nonumber\\
J_{a/b}(-\vec{q},\pm 1)&=&-\sqrt{2\pi}\sum_{J'=1}^\infty i^{J'}\sqrt{2J'+1}\left(\pm  T_{J'\pm 1}^{a/b,\mathrm{mag}}(\q)+T_{J'\pm 1}^{a/b,\mathrm{el}}(\q)\right)~,\label{eq:Jabmultipole}
\end{eqnarray}
with the transverse electric and magnetic multipole operators defined in Eq.\eqref{eq:Multipoledef}.

When substituting Eq.\eqref{eq:Jabmultipole} into Eq.\eqref{eq:T3abint},
$J'=J$ is required to conserve total angular momentum since $\phi_{i,f}$ are spinless, and parity conservation further requires that only the combinations $(a,\mathrm{el})\otimes(b,\mathrm{mag})$ and $(a,\mathrm{mag})\otimes(b,\mathrm{el})$ survive (see the discussion below Eq.\eqref{eq:Amultipole}). Since the multipole operators are irreducible tensors of the rotational group, we may apply WET,
\begin{equation}
\langle 0,0|T_{J+1}\otimes T_{J-1}'|0,0\rangle =\langle 0,0|T_{J-1}\otimes T_{J+1}'|0,0\rangle=-\langle 0,0|T_{J0}\otimes T_{J0}'|0,0\rangle
\end{equation} 
to convert the $m_J$ in all the multipole operators to 0. This leads to the following expression,
\begin{eqnarray}
iT_3^{ab}(\nu,Q^2)&=&4\pi \frac{q_0}{\q}M\sum_{J=1}^\infty(2J+1)\langle\phi_f|\bigl\{T_{J0}^{a,\mathrm{mag}}(\q)G(M+q_0+i\varepsilon)T_{J0}^{b,\mathrm{el}}(\q)\nonumber\\
&&+T_{J0}^{a,\mathrm{el}}(\q)G(M+q_0+i\varepsilon)T_{J0}^{b,\mathrm{mag}}(\q)+T_{J0}^{b,\mathrm{mag}}(\q)G(M-q_0+i\varepsilon)T_{J0}^{a,\mathrm{el}}(\q)\nonumber\\
&&+T_{J0}^{b,\mathrm{el}}(\q)G(M-q_0+i\varepsilon)T_{J0}^{a,\mathrm{mag}}(\q)\bigr\}|\phi_i\rangle~,
\end{eqnarray}
which is a useful starting point for ab-initio calculations of $T_3^{ab}(\nu,Q^2)$. One may similarly derive the multipole expansion formula for $F_3^{ab}(\nu,Q^2)$:
\begin{align}
F_3^{ab}(\nu,Q^2)&=2\pi \frac{q_0}{\q}M\sum_{J=1}^\infty(2J+1)\sum_X\delta(q_0+M-E_X)\\
&\times\big\{\langle\phi_f|T_{J0}^{a,\mathrm{mag}}(\q)|X\rangle\langle X|T_{J0}^{b,\mathrm{el}}(\q)|\phi_i\rangle
+\langle\phi_f|T_{J0}^{a,\mathrm{el}}(\q)|X\rangle\langle X|T_{J0}^{b,\mathrm{mag}}(\q)|\phi_i\rangle\big\}.\nonumber
\end{align}

\section{\label{sec:final}Summary}

Unlike the ISB correction $\delta_\text{C}$ which was studied since the late 1950s~\cite{MacDonald:1958zz}, the nuclear structure-dependent part of the RC in superallowed $\beta$ decays only received serious attention about 30 years later~\cite{Jaus:1989dh}. The quantity $\delta_\text{NS}$ originates from the difference between the ``inner'' correction on a nucleus and on a free nucleon; however, historically the two were computed in completely unrelated paradigms. The nucleon $\gamma W$-box diagram was treated in a fully relativistic framework, while $\delta_\text{NS}$ was computed with non-relativistic nuclear models. This artificial dissection may introduce additional model dependence; indeed, when the dispersive representation of the nuclear $\gamma W$-box diagram was first introduced in Refs.\cite{Seng:2018qru,Gorchtein:2018fxl}, it became clear that the previous treatments did not properly include some the important nuclear structure corrections, which now results in an inflated theory uncertainty in $|V_{ud}|_{0^+}$. As the precision of the $V_{ud}$ determination from other systems (especially from the free neutron decay) is gradually catching up, it is timely to perform a thorough re-analysis of $\delta_\text{NS}$ with modern techniques. 

In this article we present the theoretical framework to compute $\delta_\text{NS}$ in full consistency with the existing treatment of the nucleus-independent RC $\Delta_R^V$. We start with the classical Sirlin representation that allows us to rigorously separate the structure-dependent inner correction from the full $\mathcal{O}(\alpha)$ electroweak RC to the Fermi matrix element; the former is determined by the $E_e$-dependent nuclear $\gamma W$-box diagram. With this, one could unambiguously define $\delta_\text{NS}$ as the difference between the nuclear and single-nucleon $\gamma W$-box diagram, averaged over the electron energy spectrum of the nuclear $\beta$ decay. 

The $\gamma W$-box diagram is given by the $q$-integral over the spin-independent invariant amplitudes $T_i(\nu,Q^2)$ ($i=1,2,3$) (Eqs.\eqref{eq:BoxgammaW}, \eqref{eq:gammaWa}); the contribution from the parity-odd amplitude $T_3(\nu,Q^2)$ (i.e. the ``axial'' box diagram $\Box_{\gamma W}^b(E_e)$) is dominant and is the main focus of this paper. The $E_e$-independent and leading $E_e$-dependent piece are associated with the odd and even crossing components $T_{3,\pm}$, respectively. $T_{3,+}$ solely contains the isovector EM current, while $T_{3,-}$ is more complicated. For pion and free neutron $T_{3,-}$ only receives contribution from the isoscalar EM current, but for $I=1$ nuclei the isovector EM current contributes through multi-nucleon interactions, as well (see Section \ref{sec:crossing}). 

Through a DR, we expressed $T_3$ (and hence $\Box_{\gamma W}^b$) in terms of the structure function $F_3$. The dispersive representation of the nuclear box diagram is then represented in terms of the first and second Nachtmann moment of $F_3$ (Eq.\eqref{eq:boxexpand}). For the case of free nucleon where only the energy-independent part is relevant, the first Nachtmann moment $M_{3,-}^n(1,Q^2)$ is known to good precision using neutrino scattering data or indirect lattice inputs. Given that $M_{3,-}^\text{nucl}(1,Q^2)$ continues smoothly to $M_{3,-}^n(1,Q^2)$ at the asymptotic regime, the energy-independent contribution to $\delta_\text{NS}$ thus comes from the difference $M_{3,-}^\text{nucl}(1,Q^2)-M_{3,-}^n(1,Q^2)$ at low-$Q^2$, while the leading energy-dependent contribution comes from 
$M_{3,+}^\text{nucl}(2,Q^2)$ (Eq.\eqref{eq:master}); both quantities may be computed with ab-initio methods. To facilitate such calculations, we provide a number of isospin rotation formulas (see Section \ref{sec:isospin}) that relate various charge-changing nuclear matrix elements of isospin currents to diagonal nuclear matrix elements with physical electroweak currents, and also alert the readers to possible caveats in using these relations. Finally, we introduce standard nuclear physics notations that relate the relativistic invariant amplitudes and structure functions to the non-relativistic nuclear matrix elements of the nuclear Green's function $G(\omega)=(H_0-\omega)^{-1}$ (Eqs.\eqref{eq:Tmultipole}, \eqref{eq:Wmultipole}). 

Depending on the mass number $A$, different ab-initio methods may be applied for the calculation of $\delta_\text{NS}$. For instance, for light nuclei ($A\sim 10$) Green's Function Monte Carlo~\cite{Carlson:2014vla,Gandolfi:2020pbj} and no-core shell model~\cite{Barrett:2013nh} are viable. For medium-size nuclei ($A\lesssim 40$), coupled-cluster theory~\cite{Hagen:2013nca} and nuclear lattice effective field theory~\cite{Lee:2004si,Borasoy:2006qn,Lee:2008fa,Lahde:2019npb} should be applicable. The same methods may also be used to study $\delta_\text{C}$ through the newly-proposed strategy~\cite{Seng:2022epj}. We envision new calculations in the near future to pin down all the nuclear-structure dependent corrections to $V_{ud}$ with controlled theory uncertainty, and shed new light on the Cabibbo angle anomaly and other low-energy precision tests of the Standard Model. 
 
\begin{acknowledgments}

The authors thank Michael Gennari and Joanna Sobczyk for useful discussions.
The work of C.Y.S. is supported in
part by the Deutsche Forschungsgemeinschaft (DFG, German Research
Foundation), by the NSFC through the funds provided to the Sino-German Collaborative Research Center TRR110 ``Symmetries and the Emergence of Structure in QCD'' (DFG Project-ID 196253076 - TRR 110, NSFC Grant No. 12070131001), by the U.S. Department of Energy (DOE), Office of Science, Office of Nuclear Physics, under the FRIB Theory Alliance award DE-SC0013617, and by the DOE grant DE-FG02-97ER41014. The work of M.G. is supported in part by EU Horizon 2020 research and innovation programme, STRONG-2020 project
under grant agreement No 824093, and by the Deutsche Forschungsgemeinschaft (DFG) under the grant agreement GO 2604/3-1. 
  
\end{acknowledgments}

\begin{appendix}
	
\section{\label{sec:tree}Superallowed $\beta$ decay at tree level}

In this Appendix we derive the tree-level partial width of superallowed nuclear $\beta$ decays in a fully relativistic notation.
Since the kinematics are the same as that in the semileptonic kaon decay ($K_{\ell 3}$), all the exact formula in that process are directly applicable here~\cite{Seng:2019lxf,Seng:2021boy,Seng:2021wcf,Seng:2022wcw}. The relativistic formalism makes some of the higher-order corrections, e.g. recoil corrections~\cite{Holstein:1974zf}, more transparent.

After summing up the lepton spins, the absolute square of the tree-level decay amplitude (see Eq.\eqref{eq:tree}) is given by:
\begin{equation}
|\mathfrak{M}_0|^2(y,z)=G_F^2|V_{ud}|^2\left\{f_+^2 H(+1,+1)+2f_+f_-H(+1,-1)+f_-^2H(-1,-1)\right\}
\end{equation}
where 
\begin{eqnarray}
H(+1,+1)&=&2M_i^4\left[4(1-y)(y+z-1)-4r_f+r_e(r_f+4y+3z-3)-r_e^2\right]\nonumber\\
H(+1,-1)&=&2M_i^4r_e\left[3+r_e-r_f-2y-z\right]\nonumber\\
H(-1,-1)&=&2M_i^4r_e\left[1-z+r_f-r_e\right]~.
\end{eqnarray}
Here we have defined
\begin{equation}
y=\frac{2p\cdot p_e}{M_i^2}=\frac{2E_e}{M_i}~,~z=\frac{2p\cdot p'}{M_i^2}=2\frac{E_f}{M_i}~,~r_f=\frac{M_f^2}{M_i^2}~,~r_e=\frac{m_e^2}{M_i^2}~.
\end{equation}
The tree-level decay rate is given by:
\begin{equation}
\Gamma_0=\frac{M_i}{256\pi^3}\int_{\mathcal{D}_3}dydz|\mathfrak{M}_0|^2(y,z)~,
\end{equation}
where the exact three-body phase space $\mathcal{D}_3$ is given by
\begin{equation}
a(y)-b(y)<z<a(y)+b(y)~,~2\sqrt{r_e}<y<1+r_e-r_f~,
\end{equation}
with 
\begin{equation}
a(y)=\frac{(2-y)(1+r_f+r_e-y)}{2(1+r_e-y)}~,~b(y)=\frac{\sqrt{y^2-4r_e}(1+r_e-r_f-y)}{2(1+r_e-y)}~.
\end{equation}

To examine the relative importance of the form factors, let us first approximate $f_\pm(t)\rightarrow f_\pm (0)$; in this case the $z$-integration can be exactly performed. Below we display both the exact result, and the leading power-expansion according to the following counting:
\begin{equation}
m_e\sim|\vec{p}_e|\sim E_e=\mathcal{O}(\Delta)~.
\end{equation}
The results are:
\begin{eqnarray}
\int_{a-b}^{a+b}dzH(+1,+1)&=&\frac{16M_i|\vec{p}_e|(E_m^\text{exact}-E_e)^2(5E_e m_e^2 M_i-4E_eM_i^2(2E_e-M_i)-m_e^4)}{(M_i(M_i-2E_e)+m_e^2)^2}\nonumber\\
&\approx &64E_e|\vec{p}_e|(E_m-E_e)^2=\mathcal{O}(\Delta^4)\nonumber\\
\int_{a-b}^{a+b}dzH(+1,-1)&=&\frac{16m_e^2M_i|\vec{p}_e|(E_m^\text{exact}-E_e)^2(M_i(2M_i-3E_e)+m_e^2)}{(M_i(M_i-2E_e)+m_e^2)^2}\nonumber\\
&\approx&\frac{32m_e^2|\vec{p}_e|(E_m-E_e)^2}{M_i}=\mathcal{O}(\Delta^5)\nonumber\\
\int_{a-b}^{a+b}dzH(-1,-1)&=&\frac{16m_e^2M_i|\vec{p}_e|(E_m^\text{exact}-E_e)^2(E_eM_i-m_e^2)}{(M_i(M_i-2E_e)+m_e^2)^2}\nonumber\\
&\approx&\frac{16E_em_e^2|\vec{p}_e|(E_m-E_e)^2}{M_i^2}=\mathcal{O}(\Delta^6)~,\label{eq:Habintegral}
\end{eqnarray}
where 
\begin{equation}
E_m^\text{exact}\equiv\frac{M_i^2-M_f^2+m_e^2}{2M_i}~,~E_m\equiv M_i-M_f
\end{equation}
are the exact and approximate versions of the positron end-point energy, respectively.
We see that the leading order contribution comes from $H(+1,+1)$ and furthermore, $f_-$ is ISB-suppressed. Therefore, for practical purposes it is safe to retain only the $f_+^2$ term. 
In the meantime, 
\begin{equation}
\int_{2\sqrt{r_e}}^{1+r_e-r_f}dy=\frac{2}{M_i}\int_{m_e}^{E_m^\text{exact}}dE_e\approx \frac{2}{M_i}\int_{m_e}^{E_m}dE_e~.
\end{equation}
So, 
\begin{equation}
\frac{M_i}{256\pi^3}\int_{\mathcal{D}_3}dydz|\mathfrak{M}_0|^2\approx \frac{1}{2\pi^3}G_F^2|V_{ud}|^2f_+^2(0)\int_{m_e}^{E_m}dE_e|\vec{p}_e|E_e(E_m-E_e)^2~.
\end{equation} 

The above is very a na\"{\i}ve treatment which misses some essential ingredient that could play significant roles especially for heavy nuclei. The most important one is the Coulomb interaction between the outgoing positron and the daughter nucleus, which is incorporated by including the Fermi function $F(Z_f,E_e)$~\cite{Fermi:1934hr}. There are also nuclear shape corrections (or, in other words, the $t$-dependence in $f_+(t)$) and the influence of the atomic electrons; they give rise to an effective shape correction function $S(Z_f,E_e)$ and a screening correction factor $Q(Z_f,E_e)$ in the integral~\cite{Hardy:2004id}. Finally, a recoil correction factor~\cite{Towner:2014uta}:
\begin{equation}
R(E_m)=1-\frac{3E_m}{M_i+M_f}
\end{equation}
takes into account (to the desired level of precision) both the small differences between the exact and approximate result of the $z$-integral (See Eq.\eqref{eq:Habintegral}, and between the exact and approximate upper limit of the $E_e$-integral.
To conclude, we may write:
\begin{equation}
|V_{ud}|^2\approx\frac{2\pi^3m_e^{-5}G_F^{-2}f_+^{-2}(0)\ln 2}{ft}~,\label{eq:Vudnaive}
\end{equation}
where $t$ is the half life, and 
\begin{equation}
f=m_e^{-5}R(E_m)\int_{m_e}^{E_m}dE_e|\vec{p}_e|E_e(E_m-E_e)^2F(Z_f,E_e)S(Z_f,E_e)Q(Z_f,E_e)\label{eq:statfunc}
\end{equation}
is the statistical rate function.

Eq.\eqref{eq:Vudnaive} is not the end of the story, as there are still several SM corrections at the level $10^{-2}-10^{-4}$ that must be included for a precise extraction of $V_{ud}$. They are: (1) the EWRC that are not already included in the Fermi function, and (2) The ISB correction $\delta_\mathrm{C}$ that mainly originates from the Coulomb interaction between the protons. In particular, the EWRC is usually divided into the nuclear-structure-independent piece $\Delta_R^V$ and the structure-dependent pieces $\delta_\mathrm{R}'$ and $\delta_\mathrm{NS}$. Adding all these gives the master formula in Eq.\eqref{eq:Vudmaster}.

\section{\label{sec:SU2}Useful theorems in SU(2)}

Many derivations in this paper make use of relations between matrix elements of angular momentum or isospin eigenstates. Since both of them are described by an SU(2) group, it is useful to state here some of the relevant group theory relations for the benefits of unfamiliar readers. In fact we just need two of them:
\begin{enumerate}
\item \textit{Product of irreducible tensors}: Suppose $H^{I_1}_{m_{I_1}}$ and $K^{I_2}_{m_{I_2}}$ are irreducible tensors of rank $I_1$ and $I_2$ respectively of an SU(2) group, then its product may be written as a sum of irreducible tensors:
\begin{equation}
H^{I_1}_{m_{I_1}}\otimes K^{I_2}_{m_{I_2}}=\sum_{Im_I}C_{I_1m_{I_1};I_2m_{I_2}}^{I_1I_2;Im_I}T^I_{m_I}~,\label{eq:tensorproduct}
\end{equation}
where $C_{I_1m_{I_1};I_2m_{I_2}}^{I_1I_2;Im_I}$ are Clebsch-Gordan (CG)-coefficients.
\item \textit{Wigner-Eckart theorem}: The matrix element of an irreducible tensor with respect to SU(2) eigenstates can be written as:
\begin{equation}
\langle I_2,m_{I_2}|T^I_{m_I}|I_1,m_{I_1}\rangle=C_{I_1m_{I_1};Im_I}^{I_1I;I_2m_{I_2}}\langle I_2||T^I||I_1\rangle~,
\end{equation}
where $\langle I_2||T^I||I_1\rangle$ is a reduced matrix element that is independent of the ``magnetic quantum numbers'' $m_{I_1}$, $m_I$ and $m_{I_2}$. 
\end{enumerate}

\section{\label{sec:derivecrossing}Crossing symmetry}

In this appendix we provide a detailed derivation of the crossing symmetry of $T_i(\nu,Q^2)$ that we presented in Sec.\ref{sec:crossing}. The basic idea follows directly the arguments in Appendix A of Ref.\cite{Seng:2018qru}. We restrict ourselves to $\beta^+$-decays. 

%\subsection{Time-reversal operation}

Our derivation starts from the following identity~\cite{Sakurai:2011zz}:
\begin{equation}
\langle\beta(p)|\hat{O}|\alpha(p)\rangle=\langle\tilde{\alpha}(\tilde{p})|\mathbb{T}\hat{O}^\dagger\mathbb{T}^{-1}|\tilde{\beta}(\tilde{p})\rangle~,
\end{equation}
where $\hat{O}$ is a linear operator, $\mathbb{T}$ is the time-reversal operator, and $|\tilde{\alpha}\rangle$, $|\tilde{\beta}\rangle$ are the time-reversed states of $|\alpha\rangle$, $|\beta\rangle$, with the time-reversed four-momentum $\tilde{p}^\mu=p_\mu$. Knowing that the time-reversal operation on a four-current reads:
\begin{equation}
\mathbb{T}J^\mu(\vec{x},t)\mathbb{T}^{-1}=J_\mu(\vec{x},-t)~,
\end{equation}
then suppose a generic tensor of time-ordered current product is defined as:
\begin{eqnarray}
T^{\mu\nu}(p,q)&=&\frac{1}{2}\int d^4x e^{iq\cdot x}\langle \phi_f(p)|T\{J_a^\mu(x) J_b^\nu(0)\}|\phi_i(p)\rangle\nonumber\\
&=&\left(-g^{\mu\nu}+\frac{q^\mu q^\nu}{q^2}\right)T_1(\nu,Q^2)+\frac{\hat{p}^\mu \hat{p}^\nu}{M\nu}T_2(\nu,Q^2)+\frac{i\epsilon^{\mu\nu\alpha\beta}p_\alpha q_\beta}{2M\nu}T_3(\nu,Q^2)~,\label{eq:Tmunuoriginal}
\end{eqnarray}
using the time-reversal operation and the definition of the time-ordered product, we obtain:
\begin{eqnarray}
T^{\mu\nu}(p,q)&=&\frac{1}{2}\int d^4x e^{-i\tilde{q}\cdot x}\langle \phi_i(\tilde{p})|T\{J^{a\dagger}_\mu(x) J^{b\dagger}_\nu(0)\}|\phi_f(\tilde{p})\rangle~.\label{eq:Treversal}
\end{eqnarray}
Comparing to the definition of  $T_A^{\mu\nu}(p,q)$ (see Eq.\eqref{eq:Tmunu}), we find that at the right hand side the upper Lorentz indices are lowered, the initial and final states are interchanged, the currents are conjugated, and the momenta are time-reversed, with an extra sign change $\tilde{q}$.

Next, suppose the matrix element in Eq.\eqref{eq:Treversal} satisfies the following relation:
\begin{equation}
\langle\phi_i|J^{a\dagger}_\mu J^{b\dagger}_\nu|\phi_f\rangle=-\xi\langle\phi_f|J^{a}_\mu J^{b}_\nu|\phi_i\rangle~\label{eq:crossrelation}
\end{equation}
where $\xi=\pm 1$, then we may plug this relation into Eq.\eqref{eq:Treversal}, express the right hand side in terms of invariant amplitudes, and compare it with Eq.\eqref{eq:Tmunuoriginal}. This gives the following crossing relations:
\begin{equation}
T_1(-\nu,Q^2)=-\xi T_1(\nu,Q^2)~,~T_{2,3}(-\nu,Q^2)=\xi T_{2,3}(\nu,Q^2)~.
\end{equation}
 
If both currents are neutral we have trivially $\xi=-1$; in flavor-changing processes the value of $\xi$ must be deduced case-by-case in each isospin channel.\\

%\subsection{With the isoscalar EM current}

We start with the isoscalar EM current. 
Using a general property of the CG coefficient,
\begin{equation}
C_{I,m_I-1;1,1}^{I,1;I,m_I}=-C_{I,m_I;1,-1}^{I,1;I,m_I-1}~,
\end{equation}
we can easily use WET in the isospin space to show that
\begin{equation}
\langle I,m_I|J_\mu^\mathrm{em(0)}(x)(J_\nu^W(0))_A|I,m_I-1\rangle=\langle I,m_I-1|J_\mu^\mathrm{em(0)}(x)(J_\nu^{W\dagger}(0))_A|I,m_I\rangle~,
\end{equation}
which implies $\xi^{(0)}=-1$,
regardless of the isospin of the external hadronic states.\\

The isovector component is more complicated and produces different results for pion, nucleon ($I=1/2$ systems) and $0^+$ nuclei ($I=1$ systems), which is often taken for granted in literature. Here we will study them carefully.

For the pion decay the isovector matrix element $T_3^{(1)}(\nu,Q^2)$ vanishes due to $G$-parity,
\begin{equation}
\langle \pi^0|J_\mathrm{em}^{(1)\mu}(J_W^{\dagger\nu})_A|\pi^+\rangle =0~,
\end{equation}
given that
\begin{equation} GJ_\mathrm{em}^{(1)\mu}G^{-1}=J_\mathrm{em}^{(1)\mu}~,~ G(J_\nu^{W\dagger})_AG^{-1}=-(J_\nu^{W\dagger})_A~,~G(\pi^{\pm,0})=-\pi^{\pm,0}~.
\end{equation}
This argument does not apply for the nucleon and $I=1$ nuclei, since they are not eigenstates of $G$-parity. To study crossing symmetry in these systems, we decompose the current product into $I=1$ and $I=2$ irreducible tensors by (anti)symmetrizing in the isospin indices,
\begin{equation}
J_\mathrm{em}^{(1)\mu}(J_W^{\dagger\nu})_A=-\frac{1}{2\sqrt{2}}V_{10}^\mu A_{1-1}^\nu=T^1_{-1}+T^2_{-1}~,\label{eq:product1}
\end{equation}
where
\begin{equation}
T^1_{-1}=-\frac{1}{4\sqrt{2}}\left(V_{10}^\mu A_{1-1}^\nu-V_{1-1}^\mu A_{10}^\nu\right)~,~T^2_{-1}=-\frac{1}{4\sqrt{2}}\left(V_{10}^\mu A_{1-1}^\nu+V_{1-1}^\mu A_{10}^\nu\right)~.\label{eq:T1T2}
\end{equation}
Similarly, we write,
\begin{equation}
J_\mathrm{em}^{(1)\mu}(J^\nu_{W})_A=\frac{1}{2\sqrt{2}}V_{10}^\mu A_{1+1}^\nu=T^1_{+1}-T^2_{+1}~.\label{eq:product2}
\end{equation}
The difference in the tensor coefficients between Eq.\eqref{eq:product1} and \eqref{eq:product2} can be understood from Eq.\eqref{eq:tensorproduct}.
To analyze crossing symmetry, we compare Eqs.\eqref{eq:product1} and \eqref{eq:product2}. For $I=1/2$ states the $T^2_{\pm 1}$ matrix element is zero, whereas
\begin{equation}
\langle 1/2,-1/2|T^1_{-1}|1/2,+1/2\rangle=-\langle 1/2,1/2|T^1_{+1}|1/2,-1/2\rangle~.
\end{equation} 
Together with Eq.\eqref{eq:crossrelation}, we find $\xi^{(1)}=+1$ for $I=1/2$ systems, in accord with Ref.\cite{Seng:2018qru}. 

For $I=1$ systems the matrix elements of both $T^1_{\pm 1}$ and $T^2_{\pm 1}$ are non-vanishing. In fact, using WET we find that they lead to opposite crossing symmetry between the matrix element of $J_\mathrm{em}^{(1)\mu}(J_W^{\dagger\nu})_A$ and $J_\mathrm{em}^{(1)\mu}(J_W^{\nu})_A$:
\begin{eqnarray}
\langle 1,m_I-1|T^1_{-1}|1,m_I\rangle&=&-\langle 1,m_I|T^1_{+1}|1,m_I-1\rangle\nonumber\\
\langle 1,m_I-1|T^2_{-1}|1,m_I\rangle&=&\langle 1,m_I|(-T^2_{+1})|1,m_I-1\rangle~.
\end{eqnarray}
This implies that $T_i^{(1)}(\nu,Q^2)$ for $I=1$ nuclear systems have mixed symmetry. 

\section{\label{sec:functions}Derivation of the Wick and residue contribution}

In this Appendix we provide some detail in the derivation of $\mathfrak{Re}\Box_{\gamma W}^{b,\text{Wick}}(E_e)$ and $\mathfrak{Re}\Box_{\gamma W}^{b,\text{res}}(E_e)$ defined in Eq.\eqref{eq:Wickandres}.	

\subsection{The Wick contribution}

The Wick contribution comes from replacing $\nu\rightarrow i\nu_E$ in Eq.\eqref{eq:boxmod}:
\begin{eqnarray}
\Box_{\gamma W}^{b,\mathrm{Wick}}(E_e)&=&\frac{4e^2}{Mf_+(0)}\int\frac{d^4q_E}{(2\pi)^4}\frac{M_W^2}{M_W^2+Q^2}\frac{1}{Q^2}\frac{Q^2-\nu_E^2-i\nu_E\frac{\vec{p}_e\cdot\vec{q}}{E_e}}{Q^2+2iE_e\nu_E-2\vec{p}_e\cdot\vec{q}}\nonumber\\
&&\times\int_{\nu_\text{thr}}^\infty d\nu'\left[\frac{F_{3,-}(\nu',Q^2)}{\nu^{\prime 2}+\nu_E^2}+\frac{i\nu_E F_{3,+}(\nu',Q^2)}{\nu'(\nu^{\prime 2}+\nu_E^2)}\right]~,\label{eq:Wickstart}
\end{eqnarray}
where $q_E=(\vec{q},\nu_E)$. 
Taking the real part gives:
\begin{eqnarray}
\mathfrak{Re}\Box_{\gamma W}^{b,\text{Wick}}(E_e)&=&\frac{4e^2}{Mf_+(0)}\int\frac{d^4q_E}{(2\pi)^4}\frac{M_W^2}{M_W^2+Q^2}\frac{1}{Q^2}\frac{1}{(Q^2-2\vec{p}_e\cdot\vec{q})^2+4E_e^2\nu_E^2}\times\nonumber\\
&&\int_{\nu_\text{thr}}^\infty d\nu'\left\{\frac{(Q^2-\nu_E^2)(Q^2-2\vec{p}_e\cdot\vec{q})-2\nu_E^2\vec{p}_e\cdot\vec{q}}{\nu^{\prime 2}+\nu_E^2}F_{3,-}(\nu',Q^2)\right.\nonumber\\
&&\left.+\frac{2E_e^2\nu_E^2(Q^2-\nu_E^2)+\nu_E^2\vec{p}_e\cdot\vec{q}(Q^2-2\vec{p}_e\cdot\vec{q})}{E_e\nu'(\nu^{\prime 2}+\nu_E^2)}F_{3,+}(\nu',Q^2)\right\}~.
\end{eqnarray}
We parameterize the Euclidean loop momentum as:
\begin{equation}
q_E=Q(\sin\phi_1\sin\phi_2\sin \phi_3,\sin\phi_1\sin\phi_2\cos\phi_3,\sin\phi_1\cos\phi_2,\cos\phi_1)~,
\end{equation}
so 
\begin{equation}
\int d^4q_E=\int_0^{2\pi}d\phi_3\int_0^\pi d\phi_2\int_0^\pi d\phi_1\int_0^\infty dQ Q^3 \sin^2\phi_1\sin \phi_2~.
\end{equation}
We may choose $\vec{p}_e$ to point along the third axis, i.e. $\vec{p}_e\cdot\vec{q}=E_eQ\sin\phi_1\cos\phi_2$, so the angle $\phi_3$ can be trivially integrated. This gives:
\begin{eqnarray}
&&\mathfrak{Re}\Box_{\gamma W}^{b,\text{Wick}}(E_e)\nonumber\\
&=&\frac{\alpha}{\pi^2 Mf_+(0)}\int_0^\infty dQ^2 \frac{M_W^2}{M_W^2+Q^2}\int_{-1}^{1}dx_1\int_{-1}^{1}dx_2\frac{Q^2\sqrt{1-x_1^2}}{(Q^2-2QE_e\sqrt{1-x_1^2}x_2)^2+4E_e^2Q^2x_1^2}\nonumber\\
&&\times\int_{\nu_\text{thr}}^\infty d\nu'\left\{\frac{(1-x_1^2)(Q^2-2E_eQ\sqrt{1-x_1^2}x_2)-2x_1^2E_eQ\sqrt{1-x_1^2}x_2}{\nu^{\prime 2}+Q^2x_1^2}F_{3,-}(\nu',Q^2)\right.\nonumber\\
&&\left.+\frac{2Q^2E_ex_1^2(1-x_1^2)+Q x_1^2\sqrt{1-x_1^2}x_2(Q^2-2E_eQ\sqrt{1-x_1^2}x_2)}{\nu'(\nu^{\prime 2}+Q^2x_1^2)}F_{3,+}(\nu',Q^2)\right\}~,
\end{eqnarray}
where $x_1\equiv\cos\phi_1$, $x_2\equiv\cos\phi_2$. Up to this point it is easy to check that the term attached to $F_{3,-}$ ($F_{3,+}$) is an even (odd) function of $E_e$, therefore we may split $\mathfrak{Re}\Box_{\gamma W}^{b,\mathrm{Wick}}(E_e)=\mathfrak{Re}\Box_{\gamma W}^{b,\mathrm{e,Wick}}(E_e)+\mathfrak{Re}\Box_{\gamma W}^{b,\mathrm{o,Wick}}(E_e)$ and study the two terms individually. 

We start from the even piece and perform $x_2$-integration which is elementary. This gives:
\begin{eqnarray}
&&\mathfrak{Re}\Box_{\gamma W}^{b,\text{e,Wick}}(E_e)\nonumber\\
&=&\frac{\alpha}{\pi^2Mf_+(0)}\int_0^\infty dQ^2\frac{M_W^2}{M_W^2+Q^2}\int_{\nu_\text{thr}}^\infty d\nu'\int_{-1}^{1}dx_1\frac{Q^2}{\nu^{\prime 2}+Q^2x_1^2}F_{3,-}(\nu',Q^2)\times\nonumber\\
&&\left\{-\frac{x_1}{4E_e^2}\left[\tan^{-1}\left(\frac{Q+2E_e\sqrt{1-x_1^2}}{2E_ex_1}\right)-\tan^{-1}\left(\frac{Q-2E_e\sqrt{1-x_1^2}}{2E_ex_1}\right)\right]\right.\nonumber\\
&&\left.+\frac{1}{4QE_e}\ln\left(\frac{Q^2+4E_eQ\sqrt{1-x_1^2}+4E_e^2}{Q^2-4E_eQ\sqrt{1-x_1^2}+4E_e^2}\right)\right\}\nonumber\\
&=&\frac{\alpha}{4\pi^2}\frac{1}{Mf_+(0)}\int_0^\infty \frac{dQ^2 }{QE_e}\frac{M_W^2}{M_W^2+Q^2}\int_{\nu_\text{thr}}^\infty d\nu'\left\{J\left(\frac{Q}{2E_e},\frac{\nu'}{Q}\right)-\frac{Q}{E_e}I\left(\frac{Q}{2E_e},\frac{\nu'}{Q}\right)\right\}F_{3,-}(\nu',Q^2)~,\nonumber\\
\end{eqnarray} 
where we have defined two integrals that are more complicated but still doable:
\begin{eqnarray}
I(a,b)&\equiv &\int_{-1}^{1}dx\frac{x}{b^2+x^2}\left[\tan^{-1}\left(\frac{a+\sqrt{1-x^2}}{x}\right)-\tan^{-1}\left(\frac{a-\sqrt{1-x^2}}{x}\right)\right]\nonumber\\
J(a,b)&\equiv&\int_{-1}^{1}dx\frac{1}{b^2+x^2}\ln\left(\frac{a^2+2a\sqrt{1-x^2}+1}{a^2-2a\sqrt{1-x^2}+1}\right)~.
\end{eqnarray}
Both integrals take distinct functional forms for $a^2<1$ and $a^2>1$. Their analytic results read:
\begin{equation}
I(a,b)=\Theta(1-a^2)I^<(a,b)+\Theta(a^2-1)I^>(a,b)~,~J(a,b)=\Theta(1-a^2)J^<(a,b)+\Theta(a^2-1)J^>(a,b)~,
\end{equation}
where 
\begin{equation}
I^<(a,b)=-\pi\ln\left(\frac{4b^2}{2b(b+\sqrt{b^2+1})+1-a^2}\right)~,~I^>(a,b)=-\pi\ln\left(1-\frac{1}{a^2(b+\sqrt{b^2+1})^2}\right)~,\label{eq:Ismallbig}
\end{equation}
and
\begin{equation}
J^<(a,b)=\frac{4\pi}{b}\tanh^{-1}\left(\frac{a}{b+\sqrt{b^2+1}}\right)~,~J^>(a,b)=\frac{4\pi}{b}\tanh^{-1}\left(\frac{1}{a(b+\sqrt{b^2+1})}\right)~.\label{eq:Jsmallbig}
\end{equation}
Next we proceed with the odd piece. Integrating out $x_2$ yields:
\begin{eqnarray}
\mathfrak{Re}\Box_{\gamma W}^{b,\text{o,Wick}}(E_e)&=&\frac{\alpha}{\pi^2Mf_+(0)}\int_0^\infty dQ^2\frac{M_W^2}{M_W^2+Q^2}\int_{\nu_\text{thr}}^\infty d\nu'\int_{-1}^{1}dx_1\frac{Q^2}{\nu^{\prime 2}+Q^2x_1^2}F_{3,+}(\nu',Q^2)\nonumber\\
&&\times\left\{\frac{x_1}{2E_e\nu'}\left[\tan^{-1}\left(\frac{Q+2E_e\sqrt{1-x_1^2}}{2E_ex_1}\right)-\tan^{-1}\left(\frac{Q-2E_e\sqrt{1-x_1^2}}{2E_ex_1}\right)\right]\right.\nonumber\\
&&\left.+\frac{Qx_1^2}{8E_e^2\nu'}\ln\left(\frac{Q^2+4E_eQ\sqrt{1-x_1^2}+4E_e^2}{Q^2-4E_eQ\sqrt{1-x_1^2}+4E_e^2}\right)-\frac{x_1^2\sqrt{1-x_1^2}}{E_e\nu'}\right\}~.
\end{eqnarray}
It may be cast in terms of the two integrals above, and two additional ones:
\begin{eqnarray}
\tilde{I}(a)&\equiv&\int_{-1}^{1}dxx\left[\tan^{-1}\left(\frac{a+\sqrt{1-x^2}}{x}\right)-\tan^{-1}\left(\frac{a-\sqrt{1-x^2}}{x}\right)\right]= \lim_{b\rightarrow\infty} b^2I(a,b)\nonumber\\
\tilde{J}(a)&\equiv&\int_{-1}^{1}dx\ln\left(\frac{a^2+2a\sqrt{1-x^2}+1}{a^2-2a\sqrt{1-x^2}+1}\right)= \lim_{b\rightarrow\infty}b^2J(a,b)~.
\end{eqnarray}
Their analytic expressions read:
\begin{equation}
\tilde{I}(a)=\Theta(1-a^2)\tilde{I}^<(a)+\Theta(a^2-1)\tilde{I}^>(a)~,~\tilde{J}(a)=\Theta(1-a^2)\tilde{J}^<(a)+\Theta(a^2-1)\tilde{J}^>(a)~,
\end{equation}
where 
\begin{equation}
\tilde{I}^<(a)=-\frac{\pi}{4}(a^2-2)~,~\tilde{I}^>(a)=\frac{\pi}{4a^2}~,\label{eq:Itildesmallbing}
\end{equation}
and 
\begin{equation}
\tilde{J}^<(a)=2\pi a~,~\tilde{J}^>(a)=\frac{2\pi}{a}~.\label{eq:Jtildesmallbig}
\end{equation}
So the final result of the odd piece reads:
\begin{eqnarray}
\mathfrak{Re}\Box_{\gamma W}^{b,{\mathrm{o,Wick}}}(E_e)&=&\frac{\alpha}{\pi^2}\frac{1}{Mf_+(0)}\int_0^\infty \frac{dQ^2}{E_e^2} \frac{M_W^2}{M_W^2+Q^2}\int_{\nu_\text{thr}}^\infty \frac{d\nu'}{\nu'} F_{3,+}(\nu',Q^2)\left\{\frac{E_e}{2}I\left(\frac{Q}{2E_e},\frac{\nu'}{Q}\right)\right.\nonumber\\
&&\left.+\frac{Q}{8}\left[-\frac{\nu^{\prime 2}}{Q^2}J\left(\frac{Q}{2E_e},\frac{\nu'}{Q}\right)+\tilde{J}\left(\frac{Q}{2E_e}\right)\right]-\frac{\pi E_e}{2Q^2}\left[2\nu'(\nu'-\sqrt{\nu^{\prime 2}+Q^2})+Q^2\right]\right\}~.\nonumber\\
\end{eqnarray}

\subsection{The residue contribution}

Next we discuss the residue contribution. We first write the full box diagram as:
\begin{eqnarray}
\Box_{\gamma W}^b(E_e)&=&\frac{4ie^2}{Mf_+(0)}\int\frac{d^4q}{(2\pi)^4}\frac{M_W^2}{M_W^2+Q^2}\frac{Q^2+\nu^2-\frac{\vec{p}_e\cdot\vec{q}}{E_e}\nu}{Q^2[(\nu-E_e)^2-|\vec{p}_e-\vec{q}|^2+i\varepsilon]}\nonumber\\
&&\times\int_{\nu_\text{thr}}^\infty d\nu'\left[\frac{F_{3,-}(\nu',Q^2)}{\nu^{\prime 2}-\nu^2}+\frac{\nu F_{3,+}(\nu',Q^2)}{\nu'(\nu^{\prime 2}-\nu^2)}\right]~,
\end{eqnarray}
and all the complexities come from the electron propagator with non-zero $E_e$. It gives two poles: $\nu=E_e+|\vec{p}_e-\vec{q}|-i\varepsilon$ and $\nu=E_e-|\vec{p}_e-\vec{q}|+i\varepsilon$. The first pole always stays in the fourth quadrant in the complex-$\nu$ plane, which does not affect the Wick rotation. But the second pole could stay either in the first or second quadrant, depends on the sign of $E_e-|\vec{p}_e-\vec{q}|$. When $E_e>|\vec{p}_e-\vec{q}|$, it lies within the Wick rotation contour and thus gives a residue contribution to the integral.

To evaluate the residue, we first parameterize the measure of the Minkowskian loop momentum $q$ as:
\begin{equation}
\int\frac{d^4q}{(2\pi)^4}=\int\frac{d^3q}{(2\pi)^4}\int_{-\infty}^{\infty}d\nu=\frac{1}{8\pi^3}\int_0^\infty d\q \q^2\int_{-1}^{1}dx\int_{-\infty}^{\infty}d\nu~,
\end{equation}
where $\q=|\vec{q}|$ and $x$ is the cosine of the angle between $\vec{p}_e$ and $\vec{q}$. Requiring a pole to exist in the first quadrant implies:
\begin{equation}
E_e>|\vec{p}_e-\vec{q}|=\sqrt{E_e^2-2E_e\q x+\q^2}\implies x>\frac{\q}{2E_e}~,
\end{equation}
which in turn restricts the $\q$-integral within 0 and $2E_e$. With this, we may apply the residue theorem in the $\nu$-integration to obtain:
\begin{eqnarray}
\Box_{\gamma W}^{b,\text{res}}(E_e)&=&\frac{2\alpha}{\pi Mf_+(0)}\int_0^{2E_e}d\q \q^2\int_{\frac{\q}{2E_e}}^{1}dx\frac{M_W^2}{M_W^2+Q^2}\frac{Q^2+\nu^2-\q\nu x}{Q^2}\frac{1}{|\vec{p}_e-\vec{q}|}\nonumber\\
&&\left.\times\int_{\nu_\text{thr}}^\infty d\nu'\left[\frac{F_{3,-}(\nu',Q^2)}{\nu^{\prime 2}-\nu^2}+\frac{\nu F_{3,+}(\nu',Q^2)}{\nu'(\nu^{\prime 2}-\nu^2)}\right]\right|_{\nu=E_e-|\vec{p}_e-\vec{q}|+i\varepsilon}~.
\end{eqnarray}

The next step is to re-express the $\{\q,x\}$ integral in terms of $\{Q,x\}$. Starting with:
\begin{equation}
Q^2=-\nu^2+\q^2=-(E_e-|\vec{p}_e-\vec{q}|)^2+\q^2=2E_e(-E_e+\q x+\sqrt{E_e^2-2E_e\q x+\q^2})~,
\end{equation}
we may solve for $\q$:
\begin{equation}
\q=\frac{-Q^2x+Q\sqrt{Q^2+4E_e^2(1-x^2)}}{2E_e(1-x^2)}~,\label{eq:qexpress}
\end{equation}
which also implies:
\begin{eqnarray}
|\vec{p}_e-\vec{q}|&=&\frac{2E_e^2(1-x^2)+Q^2-Qx\sqrt{Q^2+4E_e^2(1-x^2)}}{2E_e(1-x^2)}\nonumber\\
\nu&=&\frac{-Q^2+Qx\sqrt{Q^2+4E_e^2(1-x^2)}}{2E_e(1-x^2)}+i\varepsilon~.\label{eq:peqnu}
\end{eqnarray}
Meanwhile, the integral measure and limits become:
\begin{equation}
\int_0^{2E_e}d\q\int_{\frac{\q}{2E_e}}^{1}dx=\int_0^{2E_e}dQ\int_{\frac{Q}{2E_e}}^{1}dx\frac{2|\vec{p}_e-\vec{q}|}{\sqrt{Q^2+4E_e^2(1-x^2)}}~,
\end{equation}
which can be checked numerically by integrating both sides with respect to an arbitrary integrand. With this, the real part of the residue contribution reads:
\begin{eqnarray}
\mathfrak{Re}\Box_{\gamma W}^{b,\text{res}}(E_e)&=&\frac{2\alpha}{\pi Mf_+(0)}\int_0^{2E_e}dQ\int_{\frac{Q}{2E_e}}^{1}dx\frac{M_W^2}{M_W^2+Q^2}\frac{Q^2+\nu^2-\q\nu x}{Q^2}\frac{2\q^2}{\sqrt{Q^2+4E_e^2(1-x^2)}}\nonumber\\
&&\times\int_{\nu_\text{thr}}^\infty d\nu'\left[F_{3,-}(\nu',Q^2)+\frac{\nu}{\nu'}F_{3,+}(\nu',Q^2)\right]\text{Pr}\frac{1}{\nu^{\prime 2}-\nu^2}~,\label{eq:Boxresint}
\end{eqnarray}
where ``Pr'' denotes the principal-value. 
Using Eqs.\eqref{eq:qexpress} and \eqref{eq:peqnu}, one may again check that the term attached to $F_{3,-}$ ($F_{3,+}$) is an even (odd) function of $E_e$. So we may similarly split $\mathfrak{Re}\Box_{\gamma W}^{b,\mathrm{res}}(E_e)=\mathfrak{Re}\Box_{\gamma W}^{b,\mathrm{e,res}}(E_e)+\mathfrak{Re}\Box_{\gamma W}^{b,\mathrm{o,res}}(E_e)$.

The final step is to perform the $x$-integration in Eq.\eqref{eq:Boxresint}, remembering that $\q$ and $\nu$ are also functions of $x$. This is highly non-trivial, but fortunately the results can be expressed in terms of the functions we just defined in Eqs.\eqref{eq:Ismallbig}, \eqref{eq:Jsmallbig}, \eqref{eq:Itildesmallbing} and \eqref{eq:Jtildesmallbig}:
\begin{eqnarray}
\mathfrak{Re}\Box_{\gamma W}^{b,\mathrm{e,res}}(E_e)&=&\frac{\alpha}{4\pi^2}\frac{1}{Mf_+(0)}\int_0^{4E_e^2} \frac{dQ^2}{QE_e}\frac{M_W^2}{M_W^2+Q^2}\int_{\nu_\text{thr}}^\infty d\nu' F_{3,-}(\nu',Q^2)\left\{\frac{Q}{E_e}I^<\left(\frac{Q}{2E_e},\frac{\nu'}{Q}\right)\right.\nonumber\\
&&\left.-J^<\left(\frac{Q}{2E_e},\frac{\nu'}{Q}\right)+\mathfrak{Re}\left[-\frac{Q}{E_e}I^>\left(\frac{Q}{2E_e},\frac{\nu'}{Q}\right)+J^>\left(\frac{Q}{2E_e},\frac{\nu'}{Q}\right)\right]\right\}\nonumber\\
\mathfrak{Re}\Box_{\gamma W}^{b,\mathrm{o,res}}(E_e)&=&\frac{\alpha}{\pi^2}\frac{1}{Mf_+(0)}\int_0^{4E_e^2} \frac{dQ^2}{E_e^2}\frac{M_W^2}{M_W^2+Q^2}\int_{\nu_\text{thr}}^\infty \frac{d\nu'}{\nu'} F_{3,+}(\nu',Q^2)\nonumber\\
&&\times\left\{-\frac{E_e}{2}I^<\left(\frac{Q}{2E_e},\frac{\nu'}{Q}\right)+\frac{\nu^{\prime 2}}{8Q}J^<\left(\frac{Q}{2E_e},\frac{\nu'}{Q}\right)-\frac{Q}{8}\tilde{J}^<\left(\frac{Q}{2E_e}\right)\right.\nonumber\\
&&\left.+\mathfrak{Re}\left[\frac{E_e}{2}I^>\left(\frac{Q}{2E_e},\frac{\nu'}{Q}\right)-\frac{\nu^{\prime 2}}{8Q}J^>\left(\frac{Q}{2E_e},\frac{\nu'}{Q}\right)+\frac{Q}{8}\tilde{J}^>\left(\frac{Q}{2E_e}\right)\right]\right\}~.
\end{eqnarray}
These expressions can be easily understood. First of all we recall that $\mathfrak{Re}\Box_{\gamma W}^b(E_e)=\mathfrak{Re}\Box_{\gamma W}^{b,\text{Wick}}(E_e)+\mathfrak{Re}\Box_{\gamma W}^{b,\text{res}}(E_e)$, and that the integrand of $\mathfrak{Re}\Box_{\gamma W}^{b,\text{Wick}}(E_e)$ takes two distinct forms in $Q<2E_e$ and $Q>2E_e$. Since the integrand of the residue contribution survives only in $Q<2E_e$, one naturally expects it to ``make up'' the difference so that the full expression takes a unified form in both regions. We may then immediately postulate the solutions above, which can be easily verified numerically.  

A final comment: The principal-value integration in Eq.\eqref{eq:Boxresint} activates when $\nu=\nu'$ is possible, which happens at the point:
\begin{equation}
\lim_{x\rightarrow 1}\nu=E_e-\frac{Q^2}{4E_e}=\nu'\implies E=E_\text{min}\equiv\frac{\nu'+\sqrt{\nu^{\prime 2}+Q^2}}{2}~.
\end{equation}
This means, the integrand in $\mathfrak{Re}\Box_{\gamma W}^{b,\text{res}}(E_e)$ hits a singularity when $E_e=E_\text{min}$. 

\section{\label{sec:MultipoleApp}Basics of multipole expansion}

In this Appendix we summarize some basic concepts of the multipole expansion of current operators which can be found in many references, e.g. Refs.\cite{Donnelly:1975ze,Donnelly:2017aaa}.

We start from a general current operator in the momentum space:
\begin{equation}
J^\mu(\vec{q})\equiv \int d^3x e^{-i\vec{q}\cdot\vec{x}}J^\mu(\vec{x})=(\rho(\vec{q}),\vec{J}(\vec{q}))~,
\end{equation}
where $\rho(\vec{q})$ and $\vec{J}(\vec{q})$ are the charge and (spatial-) current densities respectively. For the latter, it is customary to be spanned using the three ``spherical unit vectors'' $\vec{\epsilon}_\lambda(\vec{q})$ ($\lambda=0,\pm 1$) which satisfies $\vec{\epsilon}_\lambda(\vec{q})\cdot\vec{\epsilon}^*_{\lambda'}(\vec{q})=\delta_{\lambda\lambda'}$. The $\lambda=\pm 1$ components are transverse while the $\lambda=0$ component is longitudinal, i.e.
\begin{equation}
\vec{q}\cdot\vec{\epsilon}_{\pm 1} (\vec{q})=0~,~\vec{q}\cdot\vec{\epsilon}_0(\vec{q})\neq 0~.
\end{equation}
In practice, one usually takes $\vec{q}$ to be along the z-axis, i.e. $\vec{q}=\q\hat{e}_z$ (here $\q\equiv |\vec{q}|$). In that case, the spherical unit vectors are given by:
\begin{equation}
\vec{\epsilon}_{\pm 1}(\q\hat{e}_z)=\mp \frac{\hat{e}_x\pm i\hat{e}_y}{\sqrt{2}}~,~\vec{\epsilon}_0(\q\hat{e}_z)=\hat{e}_z~.
\end{equation}
With them we can span the spatial current as:
\begin{equation}
\vec{J}(\vec{q})=\sum_{\lambda}J(\vec{q},\lambda)\vec{\epsilon}_\lambda^*. \label{eq:Jqspan}
\end{equation}

The starting point of multipole expansion is the following plane-wave expansion formula:
\begin{equation}
e^{-i\vec{q}\cdot\vec{x}}=4\pi\sum_{J=0}^\infty\sum_{m_J=-J}^{J}(-i)^Jj_J(\q r)Y_{Jm_J}(\Omega_q)Y_{Jm_J}^*(\Omega_x)~,
\end{equation}
where $r=|\vec{x}|$. With this we can express the charge density operator as:
\begin{equation}
\rho(\vec{q})=4\pi\sum_{Jm_J}(-i)^JY_{Jm_J}^*(\Omega_q)\int d^3xM_J^{m_J}(\q,\vec{x})\rho(\vec{x})~,
\end{equation}
where 
\begin{equation}
M_J^{m_J}(\q,\vec{x})\equiv j_J(\q r)Y_{Jm_J}(\Omega_x)~.
\end{equation}

For the spatial current density operators we need one more step. We first define the ``vector spherical harmonics'' as:
\begin{equation}
\vec{Y}_{JL1}^{m_J}(\Omega_x)\equiv \sum_{m_L,\lambda}C_{Lm_L;1\lambda}^{L1;Jm_J}Y_{Lm_L}(\Omega_x)\vec{\epsilon}_\lambda~,
\end{equation}
which is a rank-$J$ tensor constructed from the product of a rank-$L$ tensor ($Y_{Lm_L}$) and a rank-1 tensor ($\vec{\epsilon}_\lambda$). With this we can write:
\begin{equation}
J(\vec{q},\lambda)=4\pi\sum_{L,m_L}\sum_{Jm_J}(-i)^L Y_{Lm_L}^*(\Omega_q)C_{Lm_L;1\lambda}^{L1;Jm_J}\int d^3x\vec{M}_{JL}^{m_J}(\q,\vec{x})\cdot \vec{J}(\vec{x})~,
\end{equation}
where
\begin{equation}
\vec{M}_{JL}^{m_J}(\q,\vec{x})\equiv j_L(\q r)\vec{Y}_{JL1}^{m_J}(\Omega_x)~.
\end{equation}	

Now we are ready to define the four types of multipole operators. They are:
\begin{eqnarray}
\text{Coulomb}:~M_{Jm_J}(\q)&\equiv &\int d^3x M_J^{m_J}(\q,\vec{x})\rho(\vec{x})~,~J\geq 0\nonumber\\
\text{Longitudinal}:~L_{Jm_J}(\q)&\equiv & \int d^3x \frac{i}{\q}\nabla M_J^{m_J}(\q,\vec{x})\cdot \vec{J}(\vec{x})~,~J\geq 0\nonumber\\
\text{Transverse electric}:~T_{Jm_J}^\mathrm{el}(\q)&\equiv &\int d^3x\frac{1}{\q}(\nabla\times\vec{M}_{JJ}^{m_J}(\q,\vec{x}))\cdot\vec{J}(\vec{x})~,J\geq 1\nonumber\\
\text{Transverse magnetic}:~T_{Jm_J}^\mathrm{mag}(\q)&\equiv &\int d^3x \vec{M}_{JJ}^{m_J}(\q,\vec{x})\cdot\vec{J}(\vec{x})~,~J\geq 1~.\label{eq:Multipoledef}
\end{eqnarray}
Using the following mathematical identity:
\begin{eqnarray}
e^{-i\vec{q}\cdot\vec{x}}\vec{\epsilon}_\lambda(\vec{q})&=&\left\{ \begin{array}{ccc}
\sqrt{4\pi}\sum_{J=0}^\infty (-i)^J\sqrt{2J+1}\frac{i}{\q}\nabla M_J^0(\q,\vec{x}) & , & \lambda=0\\
-\sqrt{2\pi}\sum_{J=1}^\infty (-i)^J\sqrt{2J+1}\left(\lambda \vec{M}_{JJ}^\lambda(\q,\vec{x})-\frac{1}{\q}\nabla\times\vec{M}_{JJ}^\lambda(\q,\vec{x})\right) & , & \lambda=\pm 1
\end{array}\right.~,\nonumber\\
\end{eqnarray}
The charge and current density operators can then be expressed as:
\begin{eqnarray}
\rho(\vec{q})&=&\sqrt{4\pi}\sum_{J=0}^\infty (-i)^J\sqrt{2J+1}M_{J0}(\q)\nonumber\\
J(\vec{q},\lambda)&=&\left\{ \begin{array}{ccc}
\sqrt{4\pi}\sum_{J=0}^\infty (-i)^J\sqrt{2J+1}L_{J0}(\q) & , & \lambda=0\\
-\sqrt{2\pi}\sum_{J=1}^\infty (-i)^J\sqrt{2J+1}\left(\lambda T_{J\lambda}^\mathrm{mag}(\q)-T_{J\lambda}^\mathrm{el}(\q)\right) & , & \lambda=\pm 1
\end{array}\right.
\end{eqnarray} 
which is our desired multipole expansion formula.

One could also define the spatial Fourier transform with reversed momentum:
\begin{equation}
J^\mu(-\vec{q})\equiv \int d^3x e^{i\vec{q}\cdot\vec{x}}J^\mu(\vec{x})=(\rho(-\vec{q}),\vec{J}(-\vec{q}))~.
\end{equation}
Here we still take $\vec{q}=\q\hat{e}_z$, with all the definitions of $\vec{\epsilon}_\lambda$ the same as before. In this case the charge and current densities can also be expanded in multipole as:
\begin{eqnarray}
\rho(-\vec{q})&=&\sqrt{4\pi}\sum_{J=0}^\infty i^J\sqrt{2J+1}M_{J0}(\q)\nonumber\\
J(-\vec{q},\lambda)&=&\left\{ \begin{array}{ccc}
-\sqrt{4\pi}\sum_{J=0}^\infty i^J\sqrt{2J+1}L_{J0}(\q) & , & \lambda=0\\
-\sqrt{2\pi}\sum_{J=1}^\infty i^J\sqrt{2J+1}\left(\lambda T_{J\lambda}^\mathrm{mag}(\q)+T_{J\lambda}^\mathrm{el}(\q)\right) & , & \lambda=\pm 1
\end{array}\right.~.
\end{eqnarray} 

Next, we know that the current could be vector or axial and could have isospin structure, see Eq.\eqref{eq:vectoraxial}.
Incorporating these details, we may define the vector multipole operators:
\begin{equation}
M_{Jm_J;Im_I}^{V}(\q)~,~L_{Jm_J;Im_I}^{V}(\q)~,~T^{V,\mathrm{el}}_{Jm_J;Im_I}(\q)~,~T^{V,\mathrm{mag}}_{Jm_J;Im_I}(\q)~,
\end{equation}
and the axial multipole operators:
\begin{equation}
M_{Jm_J;Im_I}^{A}(\q)~,~L_{Jm_J;Im_I}^{A}(\q)~,~T^{A,\mathrm{el}}_{Jm_J;Im_I}(\q)~,~T^{A,\mathrm{mag}}_{Jm_J;Im_I}(\q)\label{eq:Amultipole}
\end{equation}
accordingly. Among the vector multipoles, the Coulomb, longitudinal and transverse electric operators have parity $(-1)^J$ while the transverse magnetic operator has parity $(-1)^{J+1}$. The parity of the axial multipoles are exactly the opposite. 

Finally, in most of the nuclear theory calculations it is natural to assume isospin symmetry, so the vector isospin current $V_{Im_I}^\mu$ is conserved. As a consequence, the Coulomb and longitudinal multipole operators of the vector currents are not independent as far as their matrix elements are concerned:
\begin{equation}
\langle \phi_f|L_{Jm_J;Im_I}^V(\q)|\phi_i\rangle =-\left(\frac{E_f-E_i}{\q}\right)\langle \phi_f|M_{Jm_J;Im_I}^V(\q)|\phi_i\rangle~.
\end{equation}
One may use this relation to eliminate the vector longitudinal multipoles in favor of the vector Coulomb multipoles. 

\end{appendix}

\bibliography{deltaNS_ref}

\end{document}